\documentclass[twocolumn,english,aps,prb,showpacs,amssymb,amsfonts,superscriptaddress,longbibliography]{revtex4-1}
\usepackage[T1]{fontenc}
\usepackage[latin9]{inputenc}
\usepackage{amsmath}
\usepackage{graphicx}
\usepackage{amssymb}
\usepackage{color}
\usepackage{enumerate}
\usepackage{times}

\newcommand{\beq}{\begin{equation}}
\newcommand{\eeq}{\end{equation}}
\newcommand{\bea}{\begin{eqnarray}}
\newcommand{\eea}{\end{eqnarray}}
\newcommand{\cond}{g}

\newcommand{\addedstuff}{}

\def\s{\sigma}

\def\nhat{{\hat{\mathbf{n}}}}

\def\zhat{\hat{\mathbf{z}}}
\def\ket#1{{\left|#1\right\rangle}}

\def\bk{\mathbf{k}}
\def\bq{\mathbf{q}}

\def\bj{\boldsymbol{j}}

\def\be{\begin{eqnarray}}
\def\ee{\end{eqnarray}}

\begin{document}

\title{Probing the chiral anomaly with nonlocal transport in three dimensional topological semimetals}
\author{S.A. Parameswaran}
\affiliation{Department of Physics, University of California, Berkeley, CA 94720, USA}
\affiliation{Department of Physics and Astronomy, University of California, Irvine, CA 92697, USA}

\author{T. Grover}
\affiliation{Kavli Institute for Theoretical Physics, University of California, Santa Barbara, CA 93106, USA}
\author{D. A. Abanin}
\affiliation{Perimeter Institute for Theoretical Physics, Waterloo, ON N2L 2Y5, Canada}
\affiliation{Institute for Quantum Computing, Waterloo, ON N2L 3G1, Canada}
\author{D. A. Pesin}
\affiliation{Department of Physics and Astronomy, University of Utah, Salt Lake City, UT 84112, USA}
\affiliation{Department of Physics, California Institute of Technology, Pasadena, CA 91125, USA}
\author{A. Vishwanath}
\affiliation{Department of Physics, University of California, Berkeley, CA 94720, USA}
\affiliation{Materials Science Division, Lawrence Berkeley National Laboratories, Berkeley, CA 94720}

\date{\today}
\begin{abstract}
Weyl semimetals are three-dimensional crystalline systems where pairs of bands touch at points in momentum space, termed Weyl nodes, that are characterized by a definite topological charge: the chirality. Consequently, they exhibit the Adler-Bell-Jackiw anomaly, which in this condensed matter realization implies that application of parallel electric ($\mathbf{E}$) and magnetic ($\mathbf{B}$) fields pumps electrons between nodes of opposite chirality at a rate proportional to $\mathbf{E}\cdot\mathbf{B}$.  We argue that this pumping  is measurable via nonlocal  transport experiments, in the limit of weak internode scattering.
Specifically, we show that as a consequence of the anomaly, applying a local magnetic field parallel to an injected current induces a valley imbalance that diffuses over long distances. A probe magnetic field can then convert this imbalance into a measurable voltage drop far from source and drain. Such nonlocal transport vanishes when the injected current and magnetic field are orthogonal, and therefore serves as a test of the chiral anomaly.  
We further demonstrate that a similar effect should also characterize  Dirac semimetals --- recently reported to have been observed in experiments --- where a pair of Weyl nodes coexisting at a single point in the Brillouin zone are protected by a crystal symmetry.
Since the nodes are analogous to valley degrees of freedom in semiconductors, this suggests that valley currents in three dimensional topological semimetals can be controlled using electric fields, which has potential practical `valleytronic' applications. 
\end{abstract}
\maketitle

Weyl semimetals (WSMs) are three dimensional analogs of graphene that have received much attention following a recent proposal that they may occur in a class of iridate materials \cite{PyrochloreWeyl}.  They host electronic excitations that disperse linearly from degeneracy points at which  two energy bands meet. Near these points, the electronic states are described by 
the Weyl equation, familiar from particle physics \cite{Herring, Abrikosov, VolovikBook,PyrochloreWeyl}, and possess a definite {\it chirality}. While the robustness of such two-fold band-touchings --- which require the breaking of either time reversal or inversion symmetries --- has long been known \cite{Herring, Abrikosov}, the topological aspects of WSMs were only appreciated more recently \cite{NielsenABJ, PyrochloreWeyl,TurnerReview,VolovikBook}. A Weyl node is a topological object:  depending on its chirality it acts as a source or sink of Chern flux in the Brillouin zone. Since the total Chern flux through the Brillouin zone must vanish, Weyl nodes necessarily occur in pairs of opposite chirality. This topological property of the nodes
protects a single Weyl node against opening a gap: in order to remove a Weyl band-touching, a perturbation must necessarily couple the nodes, and thus WSMs should be robust against smooth disorder that only weakly mixes nodes separated in momentum space.

  {\addedstuff
Closely related to the WSM is the {\it Dirac} semimetal\cite{YoungDSM,FangNa3BiDirac, FangCd3As2Dirac} (DSM), where a {\it pair} of Weyl nodes of opposite chirality coexist at a point in the three-dimensional Brillouin zone  --- and therefore {\it four} bands touch, rather than two. Although na\"ively it appears that this situation would be unstable against a variety of gap-opening scenarios, in certain cases the resulting gapped phases always break a crystalline point-group symmetry. Therefore, as long as such symmetries are preserved, the Dirac point\footnote{Throughout, we will be careful to distinguish between Dirac {\it points} and Weyl {\it nodes}, so that we can  refer to the two Weyl nodes at a single Dirac point.} remains stable. A convenient picture of the simplest DSM is two copies of a WSM, with each copy labeled by a different crystalline point group `isospin' index. (To avoid confusion, we  refer to the valley degree of freedom common to both cases as `pseudospin'.) DSMs are thus {\it crystalline symmetry-protected} topological semimetals, and from the preceding discussion it  should be evident that stable three-dimensional Dirac points appear in pairs that lie on axes of high crystal symmetry. Recent  photoemission\cite{Na3BiExpt1,Cd3As2Expt1,Cd3As2Expt2,Na3BiExpt2} and magnetotransport\cite{Ong_unpub} measurements appear to support the theoretical prediction\cite{FangNa3BiDirac, FangCd3As2Dirac} that the three-dimensional materials Na$_3$Bi and Cd$_3$As$_2$ host DSM phases.
 } 
 
Such robust topological phases are typically characterized by the presence of protected surface states, or by unusual electromagnetic (EM) responses. WSMs are no exception: for instance, in Ref.~\onlinecite{PyrochloreWeyl}, it was demonstrated that they possess protected chiral Fermi arc surface states. {\addedstuff Similar features are also expected for DSMs, as long as the protecting crystal symmetry remains unbroken; the resulting surface states, of course, now carry additional isospin labels. }

The unconventional bulk EM response of a single three-dimensional Weyl node is known as the Adler-Bell-Jackiw anomaly \cite{AdlerAnomaly,BellJackiwAnomaly,NielsenABJ}: simultaneous application of parallel electric and magnetic  fields (applying `$\mathbf{E}\cdot\mathbf{B}$') leads to production or depletion of charge, depending on the chirality. This is clearly incompatible with charge conservation. The appearance of a Weyl node of opposite chirality resolves this apparent contradiction, since the charge produced (depleted) at one node is accounted for by that depleted (produced) at the other, and it is clear that the total charge is conserved. However, treating the node index (hereafter, `valley', in accord with the usual semiconductor terminology) as another quantum number, it is equally clear that the valley charge is {\it not} conserved in the presence of $\mathbf{E}\cdot\mathbf{B}$: this is the chiral anomaly of the WSM. {\addedstuff For the DSM as we have mentioned, in addition to the valley pseudospin index which labels the point in the Brillouin zones where the bands touch, there is an additional two-fold `isospin' index that labels the crystalline point-group representation. The discussion of the anomaly goes through more or less unchanged for each of these isospin species.}

How can we observe this effect in experiments? 

Any proposal to detect the anomaly should be capable of distinguishing anomaly-related physics from conventional metallic behavior, and should ideally {\it vanish} in the absence of the anomaly -- {\it i.e.}, either when $\mathbf{E}\cdot\mathbf{B}=0$ or in the absence of Weyl nodes; furthermore, it should be applicable to topological semimetals realized in several different systems --- in other words, we seek a response characteristic of the phase rather than of any specific realization.

\begin{figure}
\includegraphics[width=\columnwidth]{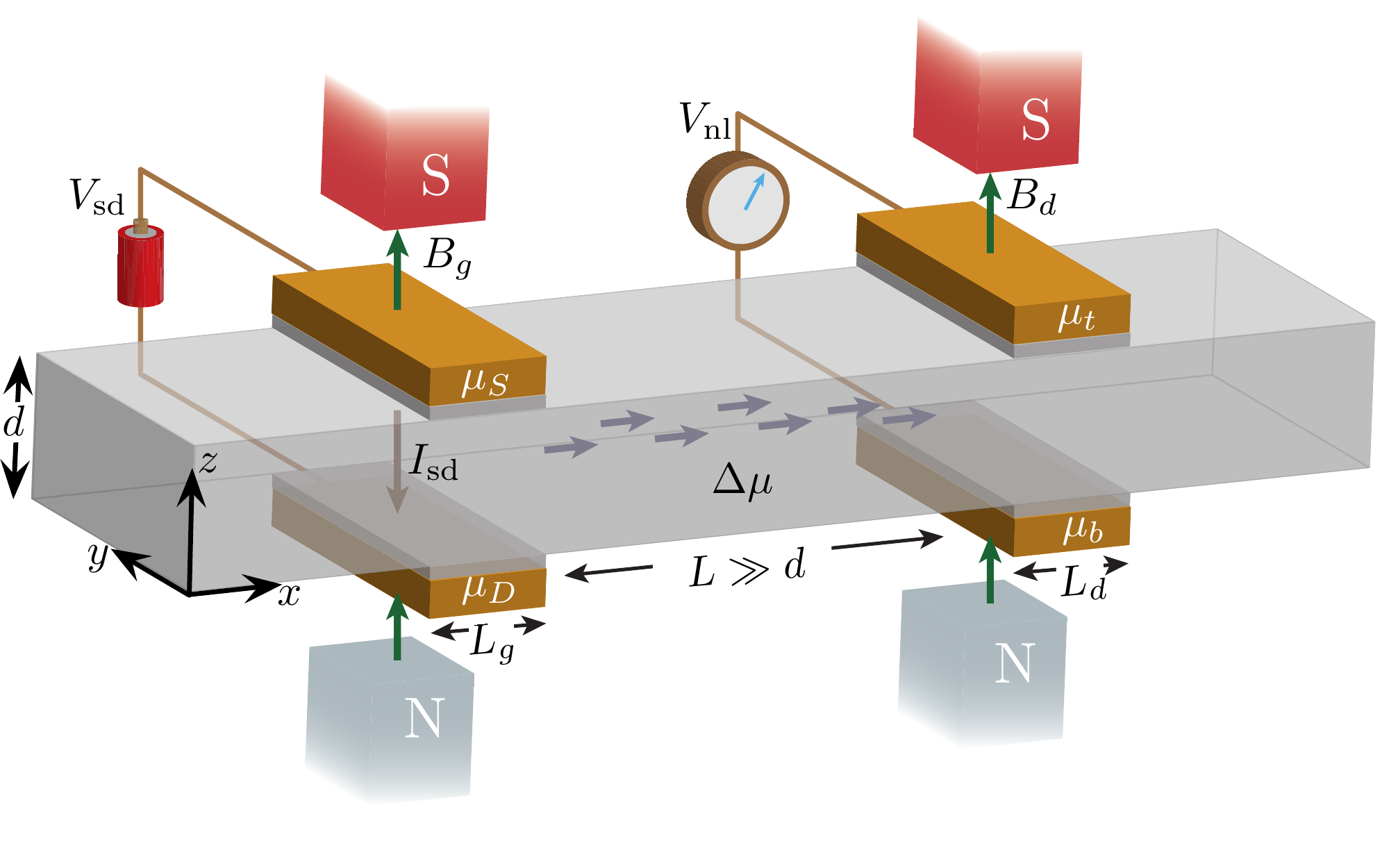}
\caption{{\bf \label{fig:transport}Nonlocal Transport Experiment.} A source-drain current $I_{\text{sd}}$ is injected into a Weyl semimetal slab of thickness $d$ via tunneling contacts of thickness $L_g$.  In the presence of a local `generation' magnetic field $B_g$, a valley imbalance $\Delta\mu$ is created via the chiral anomaly and diffuses a distance $L\gg d$ away. If a `detection' field $B_d$ is applied, the valley imbalance can be converted into a potential difference $V_{\text{nl}}$ between top and bottom contacts of size $L_d$.}
\end{figure}

  Here we show that the slow relaxation  of valley charge (characterized by an inter-node scattering time $\tau_v$, which is typically long as it involves large quasi-momentum transfer in the Weyl case, or scattering between different point-group representations in the Dirac case) results in a signature of the charge pumping in nonlocal resistance measurements (Fig.~\ref{fig:transport}). While in general sensitive to various experimental parameters, in the ``quantum'' limit  when the valley imbalance generated is limited only by relaxation at the contacts, we find that applying a voltage $V_{\text{SD}}$ at $x=0$ 
  yields a nonlocal voltage at $x$  that is determined only by intervalley relaxation in the bulk,
\be\label{eq:simpleintroresult}
|V_{\text{nl}}(x)| =  V_{\text{SD}}e^{-x/\ell_v},
\ee 
where $\ell_v =\sqrt{D \tau_v}$ is the valley relaxation length, and $D$ is the charge diffusion coefficient. In contrast, conventional Ohmic voltages decay on the scale of the sample thickness. In addition, the dependence of nonlocal response 
on field orientation (described in detail below) further reflects its origins in the anomaly. In particular, $V_{\text{nl}} =0$ when an $\mathbf{E}\cdot\mathbf{B}$ term is absent.  Thus in the idealized limit, $\ell_v\rightarrow \infty$, this gives an unambiguous signature of the anomaly.

The proposed experiment is easily sketched; for the moment, let us focus on the Weyl case. 
First, a charge current is driven across the sample in a region where a local magnetic field $B_g$ is applied. Due to the chiral anomaly, in the steady state a valley imbalance -- in the ideal case, proportional to $B_g$ --  is generated in the region where charge current flows.  As long as the inter-node scattering is weak, the valley imbalance is long-lived and can diffuse far away ({\it i.e.}, a distance of order $\ell_v$) from the region where it is generated. In the absence of a magnetic field, the valley imbalance does not couple to an electric field and is thus challenging to detect. However, such a coupling {\it does} arise when a local ``detection'' magnetic field $B_d$ is applied -- once again, a consequence of the chiral anomaly. In this case, the valley imbalance manifests itself by building up an electrical voltage across the sample. When  $B_g$ is oriented perpendicular to the source-drain current, or $B_d$ perpendicular to the direction in which the voltage drop is measured, $V_{\text{nl}}$ vanishes, reflecting the fact that the anomaly is sensitive to the angle between $\mathbf{E}$ and $\mathbf{B}$.
While this nonlocal effect bears some resemblance to the so-called `Zeeman-driven spin Hall effect' and associated transport phenomena in graphene~\cite{AbaninNonlocal, AbaninGiantNonlocality, AbaninZeemanSHE, Balakrishnan:2013fk}  and  other spin Hall materials,  the dependence on the orientation of the magnetic field is unique and is a signature of the chiral anomaly.

Before proceeding, we briefly review other recent proposals to study the anomaly in WSMs. One observation \cite{NielsenABJ}  is that there is an additional, anomaly-induced current along the magnetic field direction; the resulting  anisotropy in the magnetoconductance has been suggested \cite{AjiABJAnomaly, SonSpivak} as a signature of the anomaly. However, one might expect such anisotropy simply on symmetry grounds \cite{BurkovTopoNodal} in a conventional metal, since the magnetic field provides a preferred direction. 
Additionally, the anomaly results in a { negative} classical magnetoresistance that can be quite large\cite{SonSpivak}; however this remains a quantitative rather than a qualitative signature, and may be overwhelmed by other contributions making it challenging to detect.
 Another proposal \cite{QiPlasmons} is to realize a WSM by magnetically doping a topological insulator; the symmetry-breaking  ferromagnetic order can then be used as a probe of the underlying topological semimetal. For instance, as a consequence of the anomaly, vortex lines in the ferromagnet carry one-dimensional chiral modes, and the ferromagnetic Goldstone modes couple to charge plasmons. While striking, such features are not easy to probe, and are specific to the example studied rather than serving as a general signature of a WSM. In Ref.~\onlinecite{RanWeyl}, an anomalous Hall effect signature was discussed, related to the Weyl anomaly \cite{TurnerReview}; however this is absent for certain high symmetry crystals and needs additional information on the momentum space location of nodes to be turned into a sharp signature. Thus, existing approaches to study the topological response of WSMs stand in marked contrast to the simple transport experiment proposed here, which applies generally to all realizations of  WSMs, and furthermore satisfies the criteria outlined earlier: namely, it involves a signal that is absent for conventional [semi]metals and  can be ascribed to the presence of an $\mathbf{E}\cdot\mathbf{B}$ term by examining its dependence on the orientation of the magnetic field. Furthermore, as we demonstrate, modulo some reasonable caveats about  disorder, our results apply also to DSMs.  {\addedstuff We note that, in contrast to other transport-related predictions, our proposal involves a physical mechanism --- the diffusion of valley imbalance --- that is distinct from the transport of electric charge and is crucial to the nonlocality of the response. As a consequence of the nonlocality, we may attribute our signal  to a specific magnetoresistance mechanism (namely, the anomaly) circumventing the challenges usually involved in interpreting magnetotransport measurements.} The experiment proposed here thus serves  as an `order parameter' for  topological semimetals of both the Dirac and Weyl varieties, and is the only sharp signature proposed to date that is agnostic to the specific details of the experimental realization.

In the remainder, we first outline a simple description of transport in a WSM, and proceed to discuss the  chiral charge pumping within this formalism. Having formulated a limit where the solution is especially transparent, we demonstrate the existence of a nonlocal response and examine its behavior in various cases, before turning to a simple model of impurity scattering that permits us to provide parametric estimates of various length scales; we also demonstrate that our results remain applicable to the case of a disordered DSM. We conclude with a discussion of our results and possible extensions.

\section{Model and Transport Theory for Weyl Semimetals} 
\subsection{Transport Equations}
We begin by sketching the derivation of the transport equations relevant to the problem in the WSM case. The simplest models of WSMs have  two nodes separated in momentum space, and henceforth we specialize our discussion to this situation (the extension of our results to the case with several such pairs of nodes is straightforward; we will say more about the DSM, where there are additional subtleties owing to the coincidence of two Weyl nodes at a single Dirac point, below.) Electrons emanating from the two valleys (denoted by `right' (R)  and `left' (L), and referred to as `pseudospin') are characterized by local electrochemical potentials in each valley,
\be
\mu^{R,L}_{\text{ec}} = \mu^{R,L} + e\phi,\nonumber\ee defined as the sum of the electric potential $\phi$ and the valley chemical potential $\mu^{R,L}$. We assume that each valley has the same finite doping level, so that the density of states $\nu_{3\text{D}}$ is finite, and equal in both valleys. As a result, charge transport within a valley is characterized (at $B=0$) by a finite Drude conductivity, $\sigma$, related to the diffusion coefficient $D$  via the Einstein relation $\sigma= e^2 D\nu_{3\text{D}}$.  All the chemical potentials are measured with respect to thermal equilibrium, so that the expressions below do not include any equilibrium `magnetization' currents \footnote{For instance, for $B\neq 0$ there is always a net uniform valley current $\bj^R-\bj^L \propto \mathbf{B} $ even in equilibrium, but this is unimportant to the transport calculation and  we therefore ignore it.}.

\begin{figure}
\includegraphics[width=0.74\columnwidth]{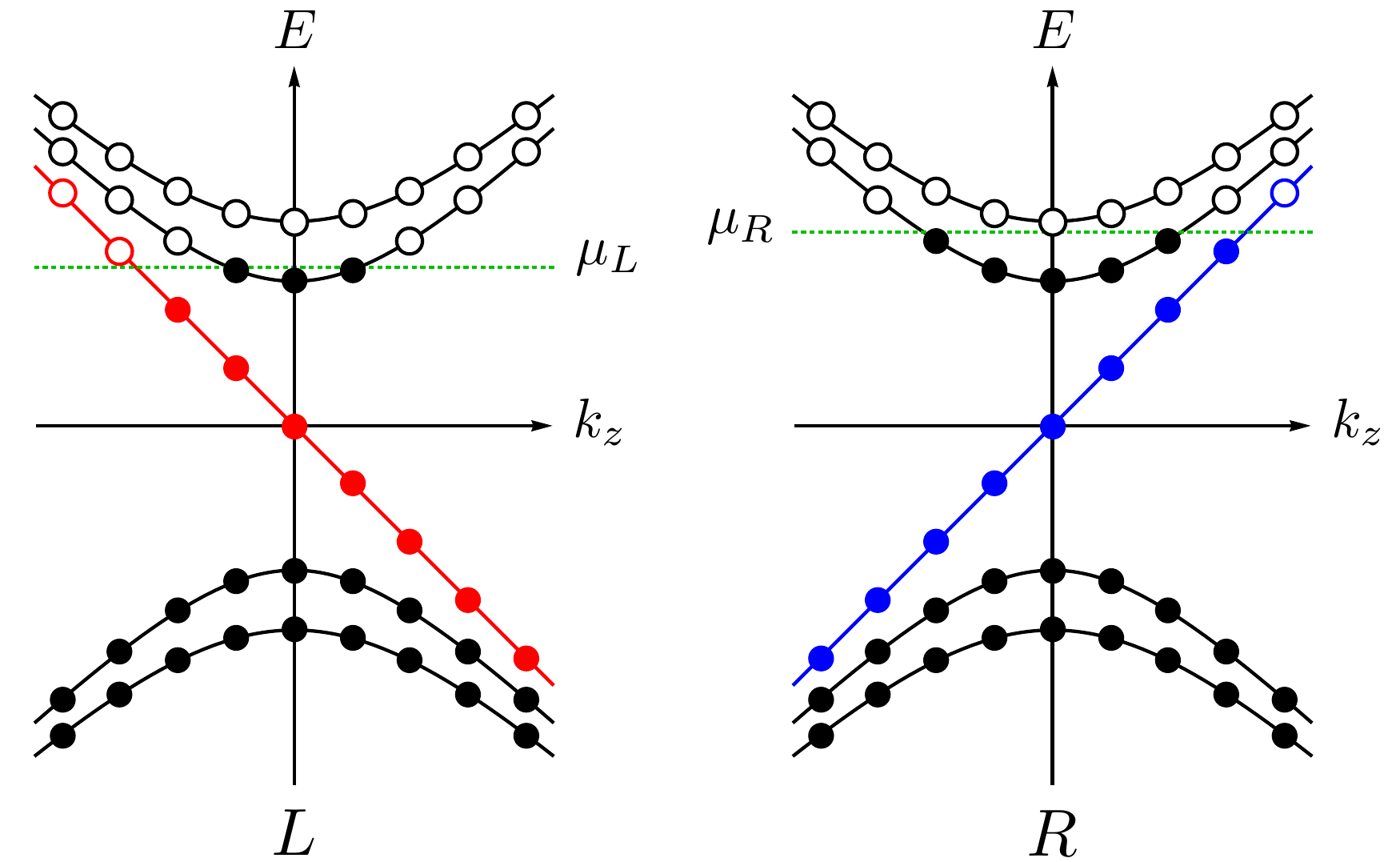}
\caption{{\bf \label{fig:WSMLLs}Landau levels at Weyl Nodes.} Filled (empty) circles denote occupied (empty) LLs. Each node has  non-chiral LLs  that disperse parabolically in the field direction (here $\zhat$) as well as a single chiral LL that disperses according to the node chirality (red, blue). A  chemical potential imbalance between the nodes leads to a net current flowing along the field, even for spatially uniform $\mu$.  }
\end{figure}

In a magnetic field $\mathbf{B} \equiv B\nhat$, the  currents in each valley can be expressed purely in terms of the potentials by solving for the Landau levels (LLs) of the Weyl nodes.  Recall that a single node \footnote{Throughout we assume an isotropic node dispersion.}  gives rise to an infinite set of LLs that disperse quadratically in the field direction, $E^{R,L}_{n}( \bk\cdot\nhat) = \hbar v_F\text{sign}(n)\sqrt{2|n|eB/\hbar c +(\bk\cdot{\nhat})^2}$ with, $n= \pm 1,\pm2,\ldots$, as well as a single ($n=0$) LL  with $E_{0}^{R,L}( \bk\cdot\nhat) = \pm \hbar v_F \bk\cdot\nhat$ that disperses linearly along the field, with a chirality set by that of the node (Fig.~\ref{fig:WSMLLs}). Each energy level is degenerate, with  $N_\Phi/A = 1/2\pi\ell_B^2$ states per unit area, where $\ell_B=({\hbar c/eB})^{1/2}$ is the magnetic length corresponding to the magnetic field $B$.  As a consequence of the chiral $n=0$ LL, electrons at a Weyl node carry a current along the field even for spatially uniform chemical potential; the total anomaly-related current is obtained by summing over all the occupied modes in this LL. In addition, in each valley there is also the conventional transport contribution due to gradients of the electrochemical potential. In our semiclassical limit, we assume this deviates only weakly from its zero-field value and is therefore well-described in terms  of the Drude conductivity $\sigma$.  These two contributions combine to give the total transport current density \cite{sonsurowka}
 \begin{equation}\label{eq:current}
 \bj^{R,L}=-\frac{\s}{e} {\boldmath{\boldsymbol{\nabla}}}\mu^{R,L}_{\text{ec}}\pm \frac{e^2 {\bf B}}{4\pi^2 \hbar^2 c}  \mu^{R,L}.
\end{equation}

These equations should be complemented by the continuity equations. In the presence of the chiral anomaly, we have \footnote{Note the extra factor of $e$ for the electrical current.}
\be
\boldsymbol{\nabla}\cdot\bj^{R,L} + {\partial_t \rho^{R,L}} = \pm \frac{e^3}{4\pi^2\hbar^2 c} \mathbf{E}\cdot\mathbf{B}
\ee
where the  $\mathbf{E}\cdot\mathbf{B} = -\mathbf{B}\cdot\boldsymbol{\nabla} \phi$ term \cite{AdlerAnomaly,BellJackiwAnomaly,NielsenABJ} is due to the anomaly, and captures the valley charge pumping.
 Using (\ref{eq:current}) and modeling inter-valley relaxation via a characteristic scattering rate by impurities $1/\tau_v$, $\partial_t \rho^{R,L} = \pm \frac{1}{2\tau_v} (\rho^R-\rho^L)$,  the steady-state continuity equations in the two valleys reduce to\begin{eqnarray}\label{eq:chem_potential}
 -\frac{\s}{e} {\nabla}^2\mu^{R,L}_{\text{ec}} \pm \frac{\beta}{e} \nhat\cdot \boldsymbol{\nabla} \mu^{R,L}_{\text{ec}} = \mp \frac{e\nu_{3\text{D}}}{2\tau_v}(\mu^R_{\text{ec}}-\mu^L_{\text{ec}}), 
\end{eqnarray}
where $\beta= \frac{1}{2\pi \ell_B^2}\frac{e^2}h$.
Note that the continuity equations depend only on the electrochemical potential, unlike the currents.

Equations (\ref{eq:current}), (\ref{eq:chem_potential}) supplemented by appropriate boundary conditions determine the charge and valley currents in the system. From this point on, we specialize to the setup illustrated in Fig.~\ref{fig:transport}, and choose coordinates in which the  $z$ direction is perpendicular to the film. 

\subsection{Boundary Conditions} 
\begin{figure}
\includegraphics[width=0.4\columnwidth]{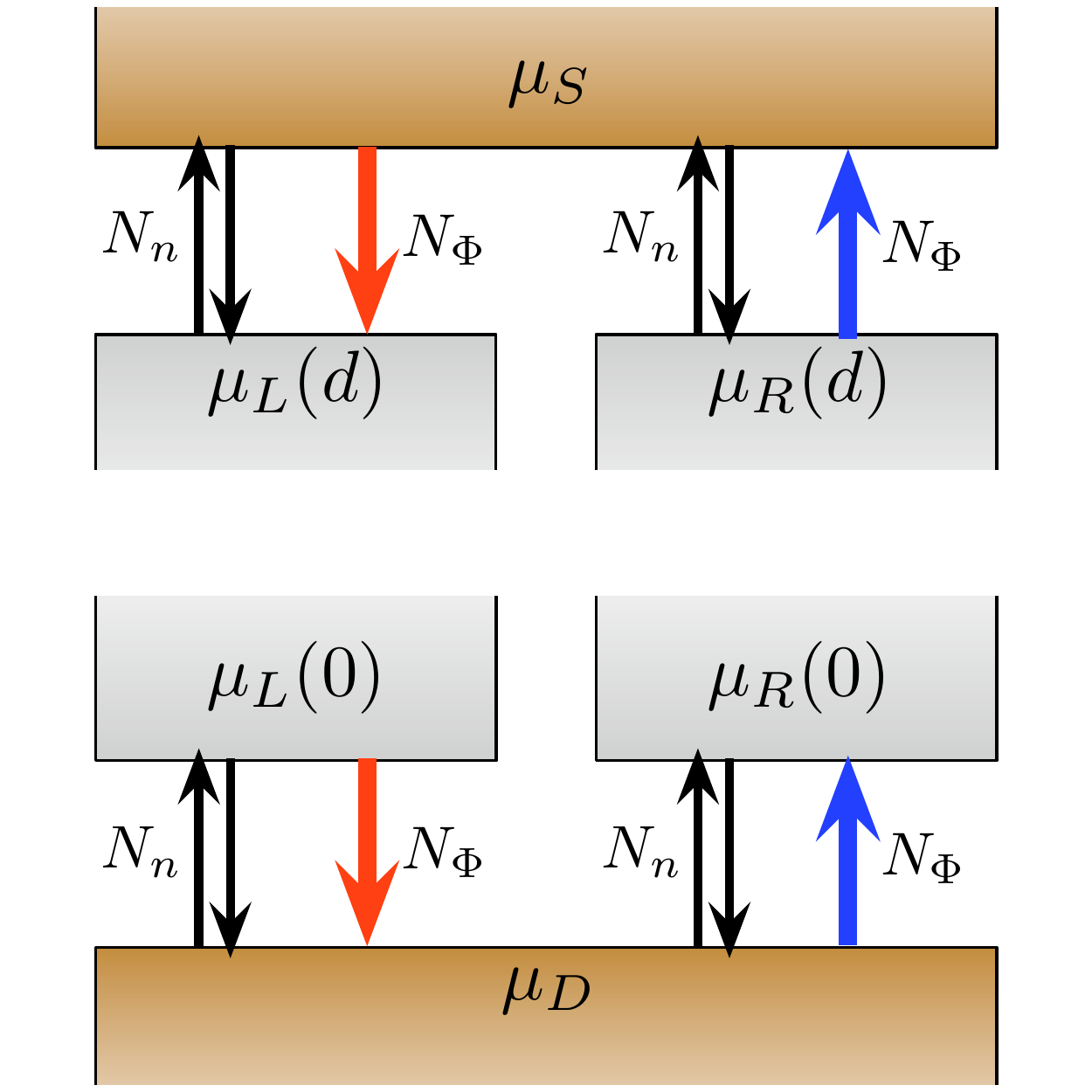}
\caption{{\bf \label{fig:contacts}Landauer Description of Contact Boundary Conditions.} Each node has $N_n\gg 1$ non-chiral `normal' transport channels in an area A (black arrows), with transmission coefficient $\mathcal{T}_i$. Since we assume that there is no intervalley scattering at the interfaces, we can simply count the number of channels and obtain a `normal' conductance per unit area $g = \frac{e^2}{h}\frac{1}{A}\sum_{i=1}^{N_n}\mathcal{T}_i$ for these channels. Here $\mathcal{T}_i$ is the transmission coefficient of channel $i$, assumed much smaller than unity for much of the paper (tunneling contacts). In addition, in each node there are also $N_\Phi$ chiral channels propagating in opposite directions in the two nodes (red/blue arrows), where $N_\Phi = A/2\pi\ell_B^2$ is the number of flux quanta threading the contact area. Applying the Landauer formalism, assuming reflectionless contacts, and assigning appropriate chemical potentials to the different channels, we obtain  boundary conditions (\ref{eq:BCtop},\ref{eq:BCbot}) for the current density.}
\end{figure}

We now establish the current boundary conditions in the presence of leads and magnetic field. 
We assume that the boundaries do not induce inter-valley scattering \footnote{Although not crucial, this simplifies the analysis.}. The boundary conditions become especially transparent when treating the interfaces in the Landauer formalism (see Fig.~\ref{fig:contacts}) and with the assumption of no intervalley scattering in the leads. 
 Let us assume that there are $N_n$ non-chiral channels in a region of area \footnote{The precise number will depend on the sample size.} $A$, each with transmission coefficient $\mathcal{T}_i$. Within the Landauer picture, these carry a current that depends on the contact conductance per unit area, $\cond = \frac{e^2}{h} \frac{1}{A}\sum_{i=1}^{N_n}\mathcal{T}_i$ and the electrochemical potential between the contact and the WSM surface. In contrast, the chiral channels in each node carry a current that depends only the electrochemical potential on one side of the interface -- {\it i.e.}, either that of the contact or of the WSM, depending on the direction. 
With these considerations, and introducing source and drain chemical potentials $\mu_{S,D}$, we find that the boundary conditions for the top surface are 
\begin{eqnarray}\label{eq:BCtop}
  j^R_z(d)&=& \frac{\cond}{e}(\mu^R_{\text{ec}}(d)-\mu_S)+ \frac{\beta}{ e} \mu^R_{\text{ec}}(d),\nonumber\\
  j^L_z(d)&=& \frac{\cond}{e}(\mu^L_{\text{ec}}(d)-\mu_S)- \frac{\beta}{e} \mu_S,
\end{eqnarray}
while on the bottom surface we have
\begin{eqnarray}\label{eq:BCbot}
  j^R_z(0)&=&\frac{\cond}{e}(\mu_D-\mu^R_{\text{ec}}(0))+ \frac{\beta}{e}\mu_D,\nonumber\\
  j^L_z(0)&=&\frac{\cond}{e}(\mu_D-\mu^L_{\text{ec}}(0))- \frac{\beta}{ e}\mu^L_{\text{ec}}(0).
\end{eqnarray}
In the above equations, it is understood that $\beta={\rm sign} (B_z)\frac{e^2}{2\pi l_B^2 h}$; we recognize this as the conductance per unit area of the chiral modes, which have $\mathcal{T}=1$. 
We will assume that in both generation and detection regions $\mathbf{B}$ lies in the $yz$ plane, inclined at angle $\theta$ from $\zhat$. In this case we have $\beta \propto \cos(\theta) = \mathbf{B}\cdot\zhat/B$. Below, we will focus on the case when the field is along $\zhat$, {\it i.e.}, $\theta=0$.

\subsection{Relaxation in Leads}
We digress briefly to discuss a subtlety that emerges when placing leads on a WSM. We wish to induce and detect disequilibrium between valley populations. However, a normal metal lead attached to a WSM has in effect a vanishingly small relaxation length for the valley degree of freedom; ideal contacts to such leads  therefore suppress  valley imbalance \footnote{ A similar problem arises with spin injection with ferromagnetic leads.}. Therefore, for a given conductivity of the WSM film, {\it tunneling} contacts to metallic leads~\cite{Rashba2000mismatch} generally  sustain a larger valley imbalance. For simplicity, we work in the thin film limit, $\cond\ll \sigma/d$, where $d$ is the thickness of the film. This condition implies that voltages are built up across the surfaces only, while inside the film fast diffusion makes the electrochemical potentials uniform across the film thickness. However, it should be kept in mind that this assumption is not necessary for nonlocal transport to arise; we will discuss the case of transparent contacts below. 

\section{Nonlocal Response}
We will work in the thin-film limit, where the various chemical and electrochemical potentials are assumed uniform across the film thickness. In this situation, averaging the continuity equations over the film thickness and using the boundary conditions~(\ref{eq:BCtop}) and ~(\ref{eq:BCbot}), we find
\begin{eqnarray}\label{eq:chempot}
  {\nabla}_\perp^2\mu^{av}_{\text{ec}}&=&\frac{2\cond+\beta}{\s d}\left(\mu^{av}_{\text{ec}}-\frac{\mu_S+\mu_D}{2}\right),\nonumber\\
  {\nabla}_\perp^2\delta\mu_{\text{ec}}&=&\frac{1}{\ell_{\text{eff}}^2}\delta\mu_{\text{ec}}+\frac{\beta}{\s d}(\mu_S-\mu_D).
\end{eqnarray}
where ${\nabla}_{\perp}^2 = \partial_{x}^2 +\partial_y^2$. We have defined the average electrochemical potential, $\mu^{av}_{\text{ec}}= \frac{1}{2d}\int_{0}^d dz (\mu^R_{\text{ec}}+\mu^L_{\text{ec}})$, and the difference between valley electrochemical potentials, $\delta \mu_{\text{ec}}=\frac{1}{d}\int_{0}^d dz(\mu^R_{\text{ec}}-\mu^L_{\text{ec}})$, and finally introduced the effective valley imbalance relaxation length, 
\begin{equation}\label{eq:ell}
  \ell_{\text{eff}}^{-2}=\frac{1}{D\tau_v}+\frac{2g+\beta}{\s d}.
\end{equation}
As discussed above, the leads induce additional intervalley relaxation, at a rate
  $ \Gamma_{\text{leads}}=\frac{2\cond+\beta}{\nu_{3\text{D}}e^2 d}.$

Equations~(\ref{eq:chempot}) allow us to analyze the valley transport in a thin WS film in the presence of external leads. The generation region is taken to consist of two massive source and drain leads of width $L_g\gg\ell_{\text{eff}}$. In this limit, they act as a valley battery, inducing in the source-drain region a valley imbalance
\begin{equation}\label{eq:Dmu}
  \delta\mu_{\text{ec}}(0)=-\frac{\beta_g\ell_{\text{eff},g}^2}{\s d}(\mu_S-\mu_D).
\end{equation}
Here, the subscript $g$ labels parameters pertaining to the generation region, and the corresponding local magnetic field $B_g$. Note that even though there is a negligible electrochemical potential drop across the film when $g\ll \sigma/d$, a valley imbalance is nevertheless generated by the preferential population of chiral modes at sample boundaries. This is essentially equivalent to the effect of the $\mathbf{E}\cdot\mathbf{B}$ term, with $\boldsymbol{\nabla}\mu_{ec}$ playing the role of the electric field.

The imbalance generated in the source-drain region diffuses over the sample, but due to intervalley scattering the imbalance decays away from the generation region with a characteristic length  $\ell_v=\sqrt{D\tau_v}$. This follows from the fact that  in the region between the `battery' and `detector' leads there are no contacts, and also we have $\mathbf{B}=0$, under which conditions we assume that the surfaces do not induce inter-valley scattering. Then the propagation length of the valley imbalance is maximal, and limited only by the weak inter-valley impurity scattering, $\ell_{\text{eff}} =\ell_v$, and so
\begin{equation} \label{eq:Dmux}
  \delta\mu_{\text{ec}}(x)=\delta\mu_{\text{ec}}(0)e^{-\frac{|x|}{\ell_v}}.
\end{equation}

To detect this imbalance far away from the generation region, one can place voltage probes in a region where  a local `detector' magnetic field, $B_d$, with the corresponding $\beta_d$, is applied. The chiral anomaly will then transform valley imbalance into charge current in the vertical direction. In order to compensate this current, a measurable voltage drop is developed between the top and bottom detecting leads. We assume that the detector is a non-invasive probe, that is, it does not alter the value of the valley imbalance it measures. This imposes a restriction on the length of the detecting leads, which will be formulated below. 

Demanding that the total current through the top and bottom contacts of the detector vanishes and using (\ref{eq:BCtop},\ref{eq:BCbot}) yields for the measured chemical potential difference between them:
\begin{equation}\label{eq:measure}
\mu_t-\mu_b=\frac{\beta_d}{2\cond_d+\beta_d}\delta\mu_{\text{ec}}(x).
\end{equation}
Using Eqs.(\ref{eq:Dmu},\ref{eq:Dmux}), we can relate the measured "nonlocal" voltage drop $V_{\text{nl}}=(\mu_t-\mu_b)/e$ to the source-drain voltage $V_{\text{SD}}={(\mu_S-\mu_D)}/{e}$. It is convenient to introduce a dimensionless coefficient $\alpha_{\text{nl}}$ that characterizes the strength of nonlocal response as a ratio of these voltages:
\be\label{eq:dimensionless}
\alpha_{\text{nl}}(x)=\frac{V_{\text{nl}}(x)}{V_{\text{SD}}}=-\frac{\beta_d}{2\cond_d+\beta_d} \frac{\beta_g\ell_{\text{eff},g}^2}{\s d}   e^{-\frac{|x|}{\ell_v}}.
\ee
This equation takes an even more transparent form if we assume that in the generation region, relaxation due to intervalley scattering can be completely neglected, and occurs solely at the leads. In this case, we can neglect the first term in the l.h.s. of Eq.(\ref{eq:ell}) for $\ell_{\text{eff},g}$, which yields
\be\label{eq:dimensionless2}
\alpha_{\text{nl}}(x) =-\frac{\beta_d}{2\cond_d+\beta_d} \frac{\beta_g}{2\cond_g+\beta_g}   e^{-\frac{|x|}{\ell_v}}.
\ee
This equation is  the central result of the paper, and gives the general dependence of the nonlocal transport on contacts, fields and intervalley relaxation in the limit when the latter is weak. Note that this condition is not unreasonable: a WSM and its attendant topological features are stable only for weak intervalley scattering, corresponding to large $\ell_v$. Furthermore, it is in this limit -- specifically, for $\ell_v\gg d$ -- that the nonlocal response dominates standard Ohmic voltages between the film surfaces. It is instructive to analyze Eq. (\ref{eq:dimensionless2}) in two  limits: 

(i) In the limit of weak generation and detection magnetic field, $\beta_g\ll \cond_g$, $\beta_d\ll \cond_d$, the nonlocal response
\be\label{eq:dimensionless3}
\alpha_{\text{nl}}(x)\approx- \frac{\beta_d}{2\cond_d} \frac{\beta_g}{2\cond_g}   e^{-\frac{|x|}{\ell_v}}.
\ee is proportional to the magnetic fields $B_g, B_d$, and changes sign if the direction of one of these fields is reversed. 
In this limit, the nonlocal voltage is inversely proportional to the conductance of the contacts $\cond_g, \cond_d$. Therefore, the nonlocal voltage is larger for tunneling contacts, as long as they continue to have a higher conductance than the chiral channels.

(ii) In the opposite limit when  $\cond_d\ll \beta_d, \cond_g\ll \beta_g$, it takes a remarkably simple form, $\alpha_{\text{nl}}(x) \approx- e^{-\frac{|x|}{\ell_v}}$,
  equivalent to that quoted in Eq.~(\ref{eq:simpleintroresult}).
and depends neither on the properties of the contacts, nor on the magnitude of the generating and detecting magnetic fields (although of course a directional dependence remains). This is the ``quantum'' limit, when the generation of valley imbalance is limited at the relaxation by the contact itself. Note that $\beta =\left({\lambda_F}/{\ell_B}\right)^2 {e^2}/{2\pi \lambda_F^2h} $, where $\lambda_F$ is the Fermi wavevector in the node, and  $e^2/h\lambda_F^2 \propto g_{\text{ideal}}$, the maximal (`Sharvin') contact conductance per unit area. In order that the semiclassical limit holds, we wish to have many filled LLs below the Fermi surface, which assumes that $\lambda_F/\ell_B$ is small, requiring $\beta\ll\cond_{\text{ideal}}$. However, since tunneling contacts  have $\cond \ll \cond_{\text{ideal}}$, this is not too restrictive, and we expect that the ``quantum'' limit can indeed be reached in experiments.

Finally, we revisit our assumption that the detector does not alter the value of the valley imbalance it measures; this constrains detector size, as follows. The detector is essentially a shunt connecting two valleys, allowing valley current to `leak' at a  rate $\Gamma_{\text{leads}}$. The detector is non-invasive if this leakage current is much smaller than the total valley current flowing under the detector, $j_v\sim D\delta\mu/\ell_v$. Comparing the latter to leakage current, $j_{\text{leak}}\sim \Gamma_{\text{leads}}L_d\delta\mu$, we obtain that the detector size, $  L_d\ll\frac{\s/d}{2\cond_d+\beta_d}\frac{d^2}{\ell_v}. $
This condition is quite weak, since $d\ll L_d\ll \ell_v$ can be satisfied for large enough $\s/d(2\cond_d+\beta_d)$, which is well within the tunneling contact/thin-film limit discussed here.
{\addedstuff
\subsection{Extension to Dirac Semimetal}
The situation in the case of a Dirac semimetal is slightly more involved; this is because at each Dirac point there is a pair of Weyl nodes of opposite chirality distinguished by the point-group index or `isospin'. Thus, for each isospin one obtains a scenario similar to that described above, with  valley imbalance having an opposite sign for the two isospin species.  In order for this simple picture to hold, we must make two crucial approximations. The first is to ignore higher-order terms in the dispersion, which lead to mixing of chiralities; the second is that we assume that the isospin relaxation time $\tau_i$ due to impurity scattering is much larger than the valley relaxation time, in order that we may treat isospin as a `good' quantum number over the length- and time-scales relevant to our experiment. We will discuss both these approximations in the next section.
Note also that the relevant time scale for valley relaxation is the {\it shorter} of $\tau_v$ and $\tau_i$, since strong scattering between isospin species will relax the valley imbalance. This follows because in any situation where at a given valley the population of one isospin species increases due to the anomaly, the population of the other isospin {\it decreases} as the anomalous contribution has the opposite sign for the two.
 With this caveat, the rest of the argument goes through identically, and one obtains a similar nonlocal transport signature as (\ref{eq:simpleintroresult}, \ref{eq:dimensionless}-\ref{eq:dimensionless3}) with $\ell_v$ replaced by $\text{min}(\sqrt{D\tau_v}, \sqrt{D\tau_i})$. In the next section, we will discuss estimates for the relevant timescales using a simple disorder model.

\section{Relaxation Processes in Disordered Dirac/Weyl Semimetals}
\begin{figure}
\includegraphics[width=0.7\columnwidth]{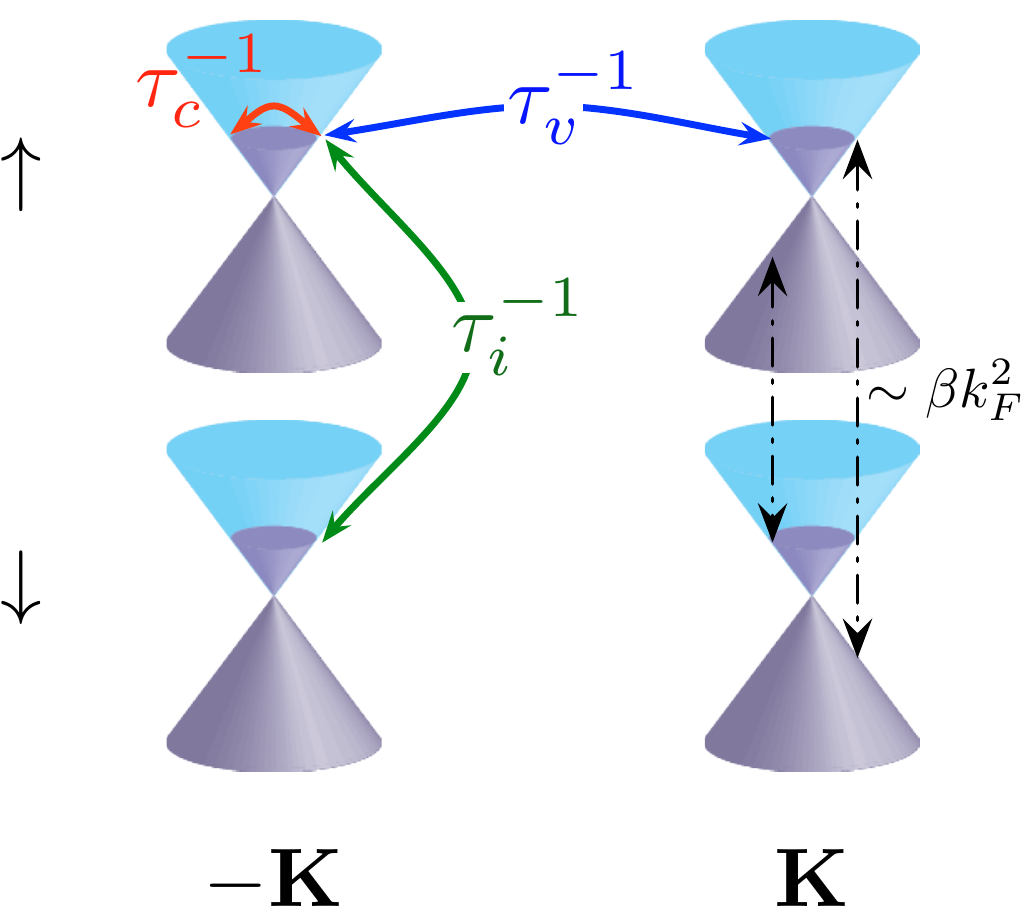}
\caption{\addedstuff {\bf \label{fig:scattering}Charge, Valley and Isospin Relaxation and Mixing in Topological Semimetals.} We depict the different  relaxation processes in disordered topological semimetals. Charge is relaxed by processes that scatter within a single Weyl node ($\tau_c^{-1}$); valley pseudospin (indicated by node position, $\pm\mathbf{K})$ is relaxed by processes involving large momentum transfer, which for screened impurities leads to strong suppression relative to charge relaxation, $\tau_v^{-1} \sim \tau_c^{-1} (k_F/2K)^4$; and in the Dirac case,  the point-group isospin (indicated by $\uparrow,\downarrow$) is relaxed by scattering between the two Weyl nodes at a single Dirac point, which are shown separately for clarity, but are in fact degenerate. Finally, curvature terms in the Dirac case mix electron-like (hole-like) isospin-$\uparrow$ states with hole-like (electron-like) isospin-$\downarrow$ states with a strength $\sim \beta k_F^2$. Note that in fact the isospin relaxation occurs by a combination of this mixing and charge relaxation at a single node, yielding $\tau_i^{-1} \sim \tau_c^{-1}(\beta k_F^2/\epsilon_F)^2$.}
\end{figure}
\subsection{Charge and Valley Relaxation in WSM}
In this section, we use a perturbative treatment of disorder to estimate the characteristic rate of two scattering processes relevant to a WSM (see Fig.~\ref{fig:scattering}):
\begin{enumerate}[(i)]
\item quasiparticle scattering at a single Weyl node, which relaxes charge imbalance at a rate  $\tau_c^{-1}$;
\item inter-valley ({\it i.e.} `pseudospin-flip' scattering, which relaxes valley imbalance at a rate  $\tau_v^{-1}$; and
\end{enumerate}
(We will discuss extensions to the Dirac case, where there is an additional issue of isospin relaxation, below.)

In order that we can distinguish the nonlocal and Ohmic responses, we require that  $d\ll \sqrt{D\tau_v}$. Furthermore, to define a local diffusive charge conductivity $\sigma$ while treating the valley imbalance as a slowly relaxing quantity, we need $\tau_c\ll \tau_v$. When both these criteria are satisfied, the nonlocal response can be clearly distinguished and is then a measure of the anomaly in the WSM.  In the DSM, we additionally require that all length scales are small compared to $\sqrt{D\tau_i}$, so that our assumption of treating the two isospin species as independent is reasonable over scales at which we measure unambiguously nonlocal effects.

We consider scattering from impurities randomly distributed with average density $n_{\text{imp}}$, each of which we shall assume is modeled by a smooth central potential $v(r)$. We will assume that (in the absence of external fields) each node is doped so that the chemical potential is away from the nodal point; we therefore take the (equilibrium) Fermi level in each node to be $\epsilon_F = \hbar v_F k_F$. We shall assume furthermore that the Fourier transform of the impurity potential takes the form 
\be
v(\bq) = \frac{v_0}{q^2+ k_{sc}^2},\ee
 where $k_{sc} = 2\pi/\lambda_{sc}$ is the characteristic  screening wavevector. We will take the two nodes to be separated by a wavevector $\mathbf{K}$. Since the screening is due to the density of electrons at a single node (characterized by $k_F$) it is reasonable to assume that $k_{sc} \sim k_{F}\ll |\mathbf{K}|$.

These minimal assumptions are sufficient to estimate the rate of relaxation of charge and valley charge and therefore are the only ones necessary for the WSM. Again, the DSM case has additional subtleties as we elucidate  below.

Using Fermi's Golden rule and averaging over disorder, we may estimate the relaxation time for quasiparticles on the Fermi surface as
\be
\tau_c^{-1} = \frac{\nu(\epsilon_F)n_{\text{imp}} }{\pi\hbar} \int \frac{d\hat{\bk}'}{4\pi} \left|v(k_F\hat{\bk}-k_F\hat{\bk}')\right|^2
\ee
where $\nu(\epsilon_F) = \epsilon_F^2/(\hbar v_F)^3$ is the density of states at the Fermi level in each node and the integral is over all possible  angular coordinates of the final state on the Fermi sphere. Using the form of the impurity potential described above and the fact that $k_F\sim k_{\text{sc}}$ to approximate the angular integration, we find
\be
\tau_c^{-1} \approx \frac{\nu(\epsilon_F)n_{\text{imp}} }{\pi\hbar} |v(0)|^2
\ee
In contrast, the charge relaxation involves a large momentum transfer as it mixes the two valleys. Parametrizing the initial and final momenta as $\bk_i = \mathbf{K} +\bk$,  $\bk_f=-\mathbf{K}+\bk'$, we have 
\be
\tau_v^{-1} = \frac{\nu(\epsilon_F)n_{\text{imp}}}{\pi\hbar} \int \frac{d\hat{\bk}'}{4\pi} \left|v(\mathbf{2K}+k_F(\hat{\bk}-\hat{\bk}'))\right|^2
\ee
Since we have $k_F\ll |\mathbf{K}|$, it is reasonable approximate this via
\be
\tau_v^{-1}  \approx \frac{\nu(\epsilon_F)n_{\text{imp}}}{\pi\hbar} |v(2\mathbf{K})|^2
\ee
Therefore, we see that the ratio of the charge and valley relaxation times is given by
\be\label{eq:chargevsvalley}
\frac{\tau_c}{\tau_v} \approx \frac{|v(2\mathbf{K})|^2}{|v(0)|^2} \approx \left(\frac{k_{sc}}{2K}\right)^4 \sim \left(\frac{k_{F}}{2K}\right)^4
\ee
where we used $|\mathbf{K}|\sim1/a\gg k_{F}$, with $a$ the lattice spacing. Thus, as long as the doping of a node, parametrized by $k_F$, is small compared to the nodal separation -- which is the criteria that the nodes are clearly resolved -- then the charge relaxation occurs on a parametrically shorter time scale than the intervalley scattering. Note that we have been a little cavalier in computing the relaxation rate rather than the transport lifetime (which differs by angular factors in the integration over final momenta), but in the limits of interest to us this distinction is negligible.

\subsection{Isospin Mixing and Relaxation in the DSM}
We now discuss the extension of our model of disorder to the DSM, where in addition to the charge and valley relaxation rates above, a crucial new quantity must be determined: the rate of isospin relaxation at a single Dirac point, due to mixing between the two Weyl nodes.  This occurs both due to weak chirality-mixing perturbations as well as impurity scattering; we denote the relaxation rate due to the latter by $\tau_{i}^{-1}$ (see Fig.~\ref{fig:scattering}). We must also make some further assumptions about the disorder. In order that our model serves as a reasonable one for estimating the { isospin} relaxation in the DSM, we require that the characteristic length scale of the potential $v(r)$ (parametrized, for instance, by the impurity screening length) is large compared to that of the crystalline unit cell, so that the precise position of the impurity within the unit cell is unimportant \footnote{In the opposite limit, even a spherically symmetric impurity might lead to mixing between isospins, as the placement of the impurity away from the symmetry center  would strongly break the crystalline point-group symmetry.}.
We defer a detailed treatment of disorder in the DSM to future work \cite{PPAV-unpub}, and for now simply sketch the argument for why the rate for isospin changing scattering processes is small. To do so, we must delve into a few more details of the DSM than we have thus far. 

The simplest model\cite{FangNa3BiDirac, FangCd3As2Dirac} of DSMs, that apply to both the cases of experimental interest, is to consider $S$ and $P$ band electrons with strong spin-orbit coupling. After incorporating the crystal field splittings allowed by the given point group symmetry, one obtains a minimal $4$-band $\bk\cdot\mathbf{p}$ Hamiltonian describing the $\ket{S_{\frac12}, \pm\frac12}$ and $\ket{P_{\frac32},\pm\frac32}$ bands; the four remaining bands mix and gap away from the Fermi level. Here we have chosen the axis of quantization of angular momentum to coincide with that of a crystalline point-group rotation. These bands interchange their valence/conduction character as one moves along the $\Gamma Q$ line in momentum space. Here. we denote by $Q$ the point on the zone boundary through which the rotation axis passes; in the standard Brillouin zone labeling convention, $Q=A$ for Na$_3$Bi, which has a hexagonal space group $P6_3/mmc$, and $Q=Z$ for Cd$_3$As$_2$ which has a tetragonal space group $P4_2/nmc$. (In both cases, we label the axis of symmetry $k_z$). Owing to the point-group symmetry along the $\Gamma Q$ line, the resulting band crossing is stable. As we move away from the $\Gamma Q$ line, the  $\ket{S_{\frac12}, \frac12}, \ket{P_{\frac32}, \frac32}$ split linearly, as do the $\ket{S_{\frac12}, -\frac12}, \ket{P_{\frac32}, -\frac32}$, but matrix elements {\it between} these pairs are quadratic in the momentum measured from the node. A simple $\bk\cdot\mathbf{p}$ matrix describing a single Dirac point that incorporates these symmetries is therefore\footnote{Note that for simplicity we have ignored an inversion-breaking term that is required to properly describe Cadmium Arsenide; we defer a discussion of such details to future work.}
\be\label{eq:DSMkpmodel}
\hat{H}(\bk) = \left(\begin{array}{cccc} v_F k_z & v_F k_+ & 0 & \beta k_-^2\\ v_F k_- & -v_F k_z & \beta k_-^2 & 0 \\ 0 & \beta k_+^2 & v_F k_z & -v_Fk_- \\ \beta k_+^2&0 & -v_F k_+ &-v_Fk_z \end{array} \right), \ee
where we have defined $k_\pm= k_x\pm i k_y$ and assumed an isotropic dispersion with $\hbar=1$, and expanded about the nodal point, assumed to be $(0,0,K)$. As we see, in the absence of the quadratic `curvature' terms (i.e., when $\beta=0$), the Dirac point can be decomposed into two independent Weyl nodes of opposite chirality\footnote{Observe that in contrast to the case of vacuum Dirac fermions familiar to high-energy physicists, this decomposition is unambiguous in a crystal since electrons emanating from the two Weyl points carry distinct crystal symmetry quantum numbers; this  is also why a mass term is forbidden by symmetry.}. In the absence of this term, rotationally invariant impurities cannot scatter between the nodes. Thus, any mixing between the two isospins depends on $\beta$. Transforming to the eigenbasis of the $\beta=0$ Hamiltonian, we find
\be\label{eq:DSMkpmodelrot}
\hat{U}_\bk\hat{H}(\bk)\hat{U}_{\bk}^{-1} = \left(\begin{array}{cccc} v_Fk &0 & 0 & \beta k_-^2\\ 0 & -v_F k & \beta k_-^2 & 0 \\ 0 & \beta k_+^2 & v_F k & 0 \\ \beta k_+^2&0 & 0 &-v_Fk \end{array} \right).
\ee
We see from this that the only mixing is between electron-like (hole-like) states of isospin `up' and hole-like (electron-like) states of isospin `down'. Taking $\epsilon_F>0$  using first-order perturbation theory in $\beta$ we find that the curvature-corrected eigenstates at the Fermi level mix the chiralities:
\be\label{eq:curvaturedressedeigstates}
\ket{{\bk}, +, \uparrow}_c \approx \ket{\bk, +, \uparrow} + \frac{\beta k}{2v_F} \sin^2\theta_{\bk}e^{-2i\phi_\bk} \ket{\bk, -,\downarrow}\nonumber\\
\ket{{\bk}, +, \downarrow}_c \approx \ket{\bk, +, \downarrow} + \frac{\beta k}{2v_F} \sin^2\theta_{\bk}e^{+2i\phi_\bk} \ket{\bk, -,\uparrow}
\ee
 where we label the two chiralities with $\uparrow$, $\downarrow$, and the electron/hole nature is indicated by the $\pm$ label, and assume that $|\bk| = k_F$. In order that we may treat the electronic levels from the two Weyl nodes at a single Dirac point as approximate chirality eigenstates, we must demand that $\beta k_F/2v_F\ll 1$. We note that a useful proxy for this assumption is that it breaks down at doping levels where the energy bands show significant deviation from linear behavior.  
 
We now turn to an estimate of the relaxation time due to impurity scattering. With the assumption of $s$-wave impurities made above, it is straightforward to estimate the scattering rate due to impurities; we find, after a few elementary manipulations and upon disorder-averaging that it is given by
\be\label{eq:impscatrate}
\tau_i^{-1} &=& \frac{\nu(\epsilon_F)n_{\text{imp}} }{\pi\hbar} \int \frac{d\hat{\bk}'}{4\pi} \left|v(k_F\hat{\bk}-k_F\hat{\bk}')\right|^2 \nonumber\\ & & \,\,\,\,\,\,\,\,\,\,\,\,\,\,\,\times \left|\,_c\langle k_F \hat{\bk}, +, \uparrow | k_F\hat{\bk}', +,\downarrow\rangle_c\right|^2
\ee
Owing to the highly anisotropic nature of the matrix element, performing the angular integration in (\ref{eq:impscatrate}) is fairly complicated in the general case. However, for our purposes it suffices to estimate an upper bound on $\tau_i^{-1}$. For this, it suffices to observe that any angular dependence of the integrand can be ignored (as these only correct the numerical prefactors), and from (\ref{eq:curvaturedressedeigstates}) that the matrix element in (\ref{eq:impscatrate}) is proportional to $\beta k_F/v_F$. Thus, we find that a rough estimate of  the impurity-induced isospin relaxation rate is
\be\label{eq:isospincharge}
\tau_i^{-1} &\approx& \frac{\nu(\epsilon_F)n_{\text{imp}} }{\pi\hbar} |v(0)|^2 \left(\frac{\beta k_F}{2v_F} \right)^2 = \tau_c^{-1}  \left(\frac{\beta k_F}{2v_F} \right)^2
\ee
It may be instructive to readers to note that this isospin relaxation process bears a mathematical resemblance to the Eliot-Yafet mechanism of spin relaxation in weakly spin-orbit coupled semiconductors. We see that once again, the criterion for this rate to be small is to require that the curvature correction to the dispersion is negligible at the relevant Fermi energy.

\section{Material Candidates}
\subsection{Weyl Semimetals}
 To date, incontrovertible evidence that any material is in the WSM phase is lacking, although the experimental situation  of transport measurements in the pyrochlore iridates  \cite{WeylResistivityMaeno}  is  encouraging, particularly in Eu$_2$Ir$_2$O$_7$ under pressure \cite{EuIridateExperiments}. In addition,  other materials have been suggested as WSM candidates~\cite{XuWSM-HgCrSe,Kim, Manes,WanOsmates,WeylAHEMFM}, and theoretical proposals to engineer Weyl nodes in topological insulator/normal insulator heterostructures \cite{WeylMultiLayer,HalaszBalents,Cho} have appeared. In the absence of an explicit realization, estimating actual values of experimental parameters is challenging. While we will  provide more  estimates with more experimental input in the more immediately compelling case of the DSM below, for now we make some very general estimates that should be broadly applicable to a variety of WSM candidates.
 
To that end, we note that the doping level $x$ in a WSM can be estimated by counting the fraction of the Brillouin zone volume occupied by the Fermi sphere:
\be\label{eq:dopinglevel}
x = \frac{2\times \frac{4\pi k_F^3}{3}}{(2K_b)^3} \approx \left(\frac{k_F}{K_b}\right)^3 
\ee
where $K_b$ is the momentum scale of the BZ and we have assumed that there are only two nodes. If we assume that $K\sim \frac{K_b}{2}$, we find using (\ref{eq:chargevsvalley}) and (\ref{eq:dopinglevel}) that
\be
\frac{\tau_c}{\tau_v} \sim \left(\frac{k_F}{2K}\right)^4 \sim {x}^{4/3}
\ee
Assuming a doping level of $1\%$, we find $\tau_v \sim 500\tau_c $. Taking a conservative estimate for the mean free path, $\ell = v_F\tau_c \sim 10\,\text{nm}$, we find that the valley relaxation length, $\ell_v$ is of the order of a few microns. We therefore see that it is not too unreasonable to expect that high-mobility samples -- where the mean free path can exceed our quite conservative estimate -- may well exhibit significant, anomaly-induced nonlocal resistance over scales where it is possible to distinguish this nonlocal signature from Ohmic conductivity.


\subsection{Dirac Semimetals}
In contrast to the Weyl case where there is as yet a convincing experimental realization, there are two promising materials that appear to exhibit Dirac semimetallic behavior in three dimensions. Following predictions from density functional theory calculations, photoemission and magnetotransport studies of the three-dimensional materials Na$_3$Bi and Cd$_3$As$_2$ strongly suggest the presence of bulk Dirac points in these materials. 

Of these, the crystal structure of  Na$_3$Bi is significantly simpler and has the added benefit of preserving inversion symmetry. In contrast, Cd$_3$As$_2$ has an 80-site unit cell in its inversion-breaking low-energy crystal structure, complicating our assumption that we can ignore the exact position of the impurity within the unit cell, and invalidating the neglect of inversion-symmetry-breaking terms that can mix  chiralities at $O(k)$ rather than $O(k^2)$. A more careful treatment of disorder than that given in the preceding section is therefore necessary. In light of this, we will focus on providing estimates for various relaxation scales in the case of Na$_3$Bi.

From the photoemission data\cite{Na3BiExpt1, Na3BiExpt2} on Na$_3$Bi, we estimate that the Fermi energy (measured relative to the node) is around $\epsilon_F \lesssim 0.02\,\text{eV}$. Note that this can be adjusted over a range of about $0.1\,\text{eV}$ by doping with potassium\cite{Na3BiExpt1}. We approximate the Fermi velocity by $v_F\sim 1\, \text{eV-\AA}$, and ignore the anisotropy in the node dispersion to obtain a characteristic Fermi wavevector $k_F\sim 0.02 \,\text{\AA}^{-1}$. Using these values in conjunction with the measured momentum-space separation of the nodes\cite{Na3BiExpt2} $2K\sim 0.2\,\text{\AA}^{-1}$ and (\ref{eq:chargevsvalley}) we find
\be
\frac{\tau_c}{\tau_v} \sim \left(\frac{k_F}{2K}\right)^4 
\sim 10^{-4}.
\ee
Thus, even the conservative estimate of the electronic mean free path such as that used above for the WSM, yields a valley relaxation length $\ell_v$ of a hundred microns. It seems reasonable to expect that various approximations (such as the neglect of anisotropy in the Fermi velocity) will only effect this estimate weakly, and that $\ell_v$ of order of tens of microns should be quite feasible, as in the WSM example above.

The curvature term is somewhat trickier to estimate. If we assume that the curvature emerges due to an underlying parabolic dispersion $\epsilon \sim k^2/2m^*$ to make a rough guess for $\beta \sim 1/2m^* \approx \left(\frac{m}{m^*}\right)\times 3.8\, \text{eV-\AA}^2$, we can estimate 
\be
\frac{\beta k_F}{2v_F } \sim 0.04\left(\frac{m}{m^*}\right) 
\ee
Since we anticipate $m^*\gtrsim m$, we see that the curvature correction is relatively small; if we set $m=m^*$ and simply use (\ref{eq:isospincharge}) to estimate the isospin relaxation time, then we find
\be
\tau_i \sim 10^3 \tau_c 
\ee
which using our estimate for the mean free path yields a relaxation length of around $10$ microns. As in the Weyl case, it is reasonable to expect that the anomaly-induced nonlocality can be clearly distinguished from Ohmic effects in samples of reasonably high mobility.
}

\section{Discussion}
We have suggested a route to studying the chiral anomaly in three dimensional topological semimetals by using it to produce and detect valley imbalance and using the slow relaxation of the latter to produce nonlocal voltage drops, which can be distinguished from more conventional Ohmic effects. Additionally, the nonlocal response is strongly dependent on the direction of applied magnetic fields, providing a means to verify its origin in the chiral anomaly. We have tried to provide the simplest description of the nonlocal response: we have assumed that the contacts dominate the relaxation at the leads and thus the process of imbalance generation and detection, and that intervalley scattering only limits the diffusion of valley imbalance away from the contacts. Furthermore, we took the contacts to be non-ideal, since metallic contacts severely constrain the generation of valley imbalance in the simple geometry proposed here. In spite of these restrictions, we find a nonlocal response that depends predominantly on parameters that can be tuned independent of the material, and no fundamental limit on the nonlocal response is apparent.

  \begin{figure}
\includegraphics[width=0.75\columnwidth]{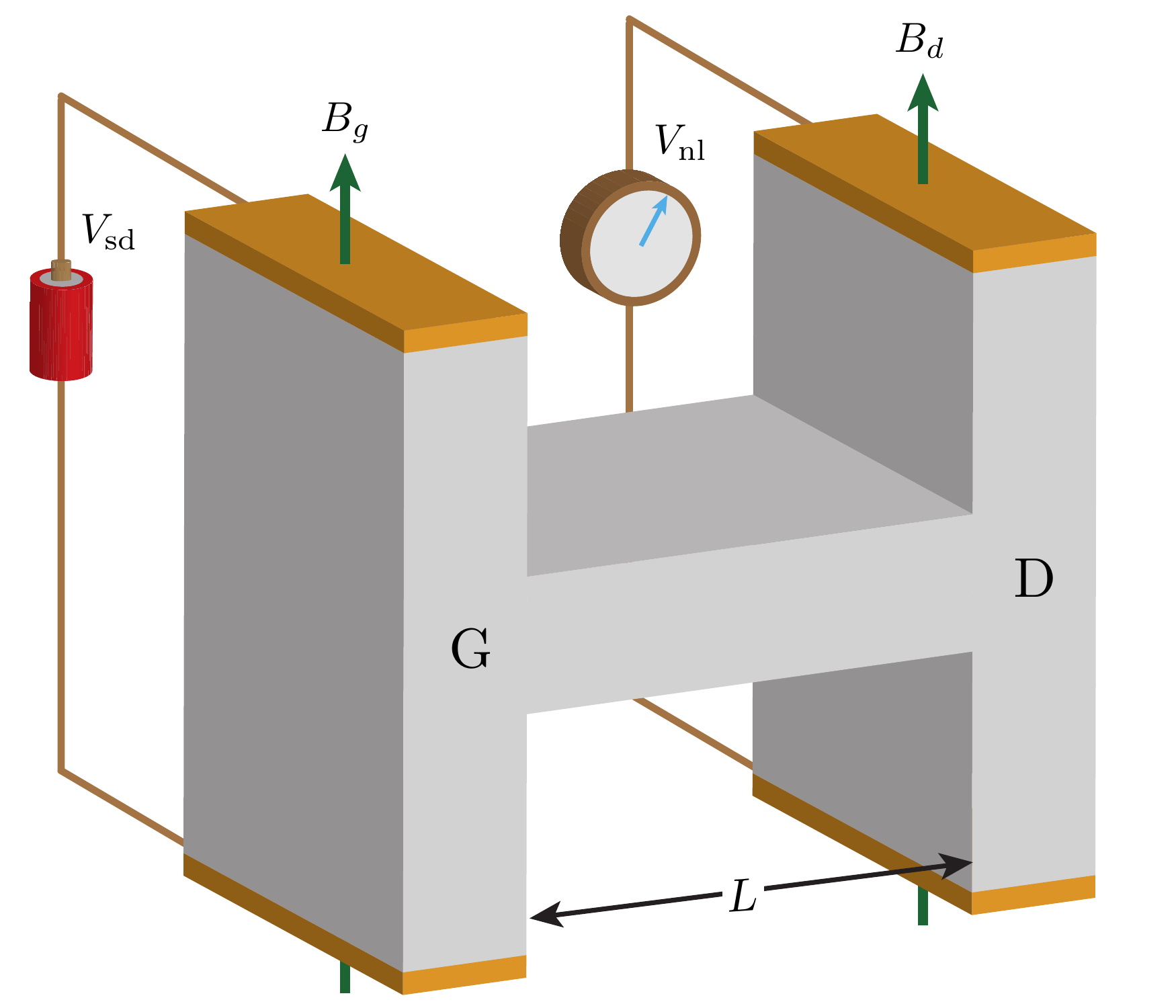}
\caption{{\bf \label{fig:HGeomFig}``H-Geometry".} An alternative setup in which the generation and detection regions are massive, so that valley relaxation at the leads is diminished, making metallic contacts feasible.}
\end{figure}

To emphasize  that the use of tunneling contacts, while important in our geometry, is not fundamental to the nonlocal transport, we note that an alternative approach would be to utilize the so-called ``H-geometry'' (Fig.~\ref{fig:HGeomFig}), frequently employed in spintronics. Here, two massive parts of the sample (generator,``G'' and detector ``D'') are connected by a narrow bridge of length $L\ll \ell_v$. Both massive parts are subject to local magnetic fields $B_g, B_d$. In addition, current is driven through region G, which leads to the generation of valley imbalance. Thus, this region acts as a ``valley battery''. The valley imbalance diffuses over the bridge to region D, where once again the chiral anomaly gives rise to a measurable voltage drop, similar to the one studied above. The H-geometry may offer some practical advantages for producing and measuring valley currents in WSMs. In particular, this geometry  increases the effective value of $d$ in the generation region. This increases the source-drain diffusion time, thereby reducing $\Gamma_{\text{leads}}$, and allows the generation of a sizable valley imbalance, limited only by intervalley impurity scattering, even with good metallic contacts.

{\addedstuff
We also constructed a simple model of scattering from screened impurities, within which we were able to provide estimates for the relevant relaxation scales applicable to the experimentally relevant case of the DSM material Na$_3$Bi. The scales obtained are within current experimental capabilities, and suggest that nonlocal magnetotransport measurements on high-mobility samples of this material may be key to unraveling its topological nature.
}

In closing we observe that since the chiral anomaly is the distinctive topological EM response of a topological semimetal, the fact that its effects have such a dramatic manifestation in relatively simple transport measurements suggests that they may be useful  in the search for both Weyl and Dirac materials in three dimensions. Similar nonlocal probes may be relevant to other cases in which topological features exist even in systems which lack a bulk gap.

\noindent{\bf Acknowledgements.-} We thank  L. Balents, J.~H. Bardarson, A. Burkov, Y.-B. Kim,  R. Ilan, N.~P. Ong and B.~Z. Spivak for useful discussions on transport,  F. de Juan, I. Kimchi, P. Dumitrescu,  N.~P. Ong and especially A. Potter for conversations on Dirac semimetals, and an anonymous referee for comments on an earlier version of this manuscript. This work was supported in part by:  the Simons Foundation (SAP); the NSF under Grant No. PHYS-1066293 and the hospitality of the Aspen Center for Physics (SAP, DAP); the Director, Office of Science, Office of Basic Energy Sciences, Materials Sciences and Engineering Division, of the U.S. Department of Energy under Contract No. DE-AC02-05CH11231 (AV); and the Institute for Quantum Information and Matter, an NSF Physics Frontiers Center with support of the Gordon and Betty Moore Foundation through Grant GBMF1250 (DAP).

\bibliography{Weyl_References}

\begin{thebibliography}{48}%
\makeatletter
\providecommand \@ifxundefined [1]{%
 \@ifx{#1\undefined}
}%
\providecommand \@ifnum [1]{%
 \ifnum #1\expandafter \@firstoftwo
 \else \expandafter \@secondoftwo
 \fi
}%
\providecommand \@ifx [1]{%
 \ifx #1\expandafter \@firstoftwo
 \else \expandafter \@secondoftwo
 \fi
}%
\providecommand \natexlab [1]{#1}%
\providecommand \enquote  [1]{``#1''}%
\providecommand \bibnamefont  [1]{#1}%
\providecommand \bibfnamefont [1]{#1}%
\providecommand \citenamefont [1]{#1}%
\providecommand \href@noop [0]{\@secondoftwo}%
\providecommand \href [0]{\begingroup \@sanitize@url \@href}%
\providecommand \@href[1]{\@@startlink{#1}\@@href}%
\providecommand \@@href[1]{\endgroup#1\@@endlink}%
\providecommand \@sanitize@url [0]{\catcode `\\12\catcode `\$12\catcode
  `\&12\catcode `\#12\catcode `\^12\catcode `\_12\catcode `\%12\relax}%
\providecommand \@@startlink[1]{}%
\providecommand \@@endlink[0]{}%
\providecommand \url  [0]{\begingroup\@sanitize@url \@url }%
\providecommand \@url [1]{\endgroup\@href {#1}{\urlprefix }}%
\providecommand \urlprefix  [0]{URL }%
\providecommand \Eprint [0]{\href }%
\providecommand \doibase [0]{http://dx.doi.org/}%
\providecommand \selectlanguage [0]{\@gobble}%
\providecommand \bibinfo  [0]{\@secondoftwo}%
\providecommand \bibfield  [0]{\@secondoftwo}%
\providecommand \translation [1]{[#1]}%
\providecommand \BibitemOpen [0]{}%
\providecommand \bibitemStop [0]{}%
\providecommand \bibitemNoStop [0]{.\EOS\space}%
\providecommand \EOS [0]{\spacefactor3000\relax}%
\providecommand \BibitemShut  [1]{\csname bibitem#1\endcsname}%
\let\auto@bib@innerbib\@empty
\bibitem [{\citenamefont {Wan}\ \emph {et~al.}(2011)\citenamefont {Wan},
  \citenamefont {Turner}, \citenamefont {Vishwanath},\ and\ \citenamefont
  {Savrasov}}]{PyrochloreWeyl}%
  \BibitemOpen
  \bibfield  {author} {\bibinfo {author} {\bibfnamefont {Xiangang}\
  \bibnamefont {Wan}}, \bibinfo {author} {\bibfnamefont {Ari~M.}\ \bibnamefont
  {Turner}}, \bibinfo {author} {\bibfnamefont {Ashvin}\ \bibnamefont
  {Vishwanath}}, \ and\ \bibinfo {author} {\bibfnamefont {Sergey~Y.}\
  \bibnamefont {Savrasov}},\ }\bibfield  {title} {\enquote {\bibinfo {title}
  {{Topological semimetal and Fermi-arc surface states in the electronic
  structure of pyrochlore iridates}},}\ }\href {\doibase
  10.1103/PhysRevB.83.205101} {\bibfield  {journal} {\bibinfo  {journal} {Phys.
  Rev. B}\ }\textbf {\bibinfo {volume} {83}},\ \bibinfo {pages} {205101}
  (\bibinfo {year} {2011})}\BibitemShut {NoStop}%
\bibitem [{\citenamefont {Herring}(1937)}]{Herring}%
  \BibitemOpen
  \bibfield  {author} {\bibinfo {author} {\bibfnamefont {Conyers}\ \bibnamefont
  {Herring}},\ }\bibfield  {title} {\enquote {\bibinfo {title} {Effect of
  time-reversal symmetry on energy bands of crystals},}\ }\href {\doibase
  10.1103/PhysRev.52.361} {\bibfield  {journal} {\bibinfo  {journal} {Phys.
  Rev.}\ }\textbf {\bibinfo {volume} {52}},\ \bibinfo {pages} {361--365}
  (\bibinfo {year} {1937})}\BibitemShut {NoStop}%
\bibitem [{\citenamefont {Abrikosov}\ and\ \citenamefont
  {Beneslavskii}(1971)}]{Abrikosov}%
  \BibitemOpen
  \bibfield  {author} {\bibinfo {author} {\bibfnamefont {A.~A.}\ \bibnamefont
  {Abrikosov}}\ and\ \bibinfo {author} {\bibfnamefont {S.~D.}\ \bibnamefont
  {Beneslavskii}},\ }\bibfield  {title} {\enquote {\bibinfo {title} {Some
  properties of gapless semiconductors of the second kind},}\ }\href@noop {}
  {\bibfield  {journal} {\bibinfo  {journal} {J. Low Temp. Phys.}\ }\textbf
  {\bibinfo {volume} {5}},\ \bibinfo {pages} {141--154} (\bibinfo {year}
  {1971})}\BibitemShut {NoStop}%
\bibitem [{\citenamefont {Volovik}(2003)}]{VolovikBook}%
  \BibitemOpen
  \bibfield  {author} {\bibinfo {author} {\bibfnamefont {G.E.}\ \bibnamefont
  {Volovik}},\ }\href@noop {} {\emph {\bibinfo {title} {{The Universe in a
  Helium Droplet}}}}\ (\bibinfo  {publisher} {Clarendon Press},\ \bibinfo
  {address} {Oxford},\ \bibinfo {year} {2003})\BibitemShut {NoStop}%
\bibitem [{\citenamefont {Nielsen}\ and\ \citenamefont
  {Ninomiya}(1983)}]{NielsenABJ}%
  \BibitemOpen
  \bibfield  {author} {\bibinfo {author} {\bibfnamefont {H.B.}\ \bibnamefont
  {Nielsen}}\ and\ \bibinfo {author} {\bibfnamefont {Masao}\ \bibnamefont
  {Ninomiya}},\ }\bibfield  {title} {\enquote {\bibinfo {title} {{The
  Adler-Bell-Jackiw anomaly and Weyl fermions in a crystal}},}\ }\href
  {\doibase 10.1016/0370-2693(83)91529-0} {\bibfield  {journal} {\bibinfo
  {journal} {Phys. Lett. B}\ }\textbf {\bibinfo {volume} {130}},\ \bibinfo
  {pages} {389 -- 396} (\bibinfo {year} {1983})}\BibitemShut {NoStop}%
\bibitem [{\citenamefont {Turner}\ and\ \citenamefont
  {Vishwanath}(2013)}]{TurnerReview}%
  \BibitemOpen
  \bibfield  {author} {\bibinfo {author} {\bibfnamefont {Ari~M.}\ \bibnamefont
  {Turner}}\ and\ \bibinfo {author} {\bibfnamefont {Ashvin}\ \bibnamefont
  {Vishwanath}},\ }\href@noop {} {\enquote {\bibinfo {title} {{Beyond Band
  Insulators: Topology of Semi-metals and Interacting Phases}},}\ } (\bibinfo
  {year} {2013}),\ \bibinfo {note} {unpublished},\ \Eprint
  {http://arxiv.org/abs/1301.0330} {arXiv:1301.0330} \BibitemShut {NoStop}%
\bibitem [{\citenamefont {Young}\ \emph {et~al.}(2012)\citenamefont {Young},
  \citenamefont {Zaheer}, \citenamefont {Teo}, \citenamefont {Kane},
  \citenamefont {Mele},\ and\ \citenamefont {Rappe}}]{YoungDSM}%
  \BibitemOpen
  \bibfield  {author} {\bibinfo {author} {\bibfnamefont {S.~M.}\ \bibnamefont
  {Young}}, \bibinfo {author} {\bibfnamefont {S.}~\bibnamefont {Zaheer}},
  \bibinfo {author} {\bibfnamefont {J.~C.~Y.}\ \bibnamefont {Teo}}, \bibinfo
  {author} {\bibfnamefont {C.~L.}\ \bibnamefont {Kane}}, \bibinfo {author}
  {\bibfnamefont {E.~J.}\ \bibnamefont {Mele}}, \ and\ \bibinfo {author}
  {\bibfnamefont {A.~M.}\ \bibnamefont {Rappe}},\ }\bibfield  {title} {\enquote
  {\bibinfo {title} {Dirac semimetal in three dimensions},}\ }\href {\doibase
  10.1103/PhysRevLett.108.140405} {\bibfield  {journal} {\bibinfo  {journal}
  {Phys. Rev. Lett.}\ }\textbf {\bibinfo {volume} {108}},\ \bibinfo {pages}
  {140405} (\bibinfo {year} {2012})}\BibitemShut {NoStop}%
\bibitem [{\citenamefont {Wang}\ \emph {et~al.}(2012)\citenamefont {Wang},
  \citenamefont {Sun}, \citenamefont {Chen}, \citenamefont {Franchini},
  \citenamefont {Xu}, \citenamefont {Weng}, \citenamefont {Dai},\ and\
  \citenamefont {Fang}}]{FangNa3BiDirac}%
  \BibitemOpen
  \bibfield  {author} {\bibinfo {author} {\bibfnamefont {Zhijun}\ \bibnamefont
  {Wang}}, \bibinfo {author} {\bibfnamefont {Yan}\ \bibnamefont {Sun}},
  \bibinfo {author} {\bibfnamefont {Xing-Qiu}\ \bibnamefont {Chen}}, \bibinfo
  {author} {\bibfnamefont {Cesare}\ \bibnamefont {Franchini}}, \bibinfo
  {author} {\bibfnamefont {Gang}\ \bibnamefont {Xu}}, \bibinfo {author}
  {\bibfnamefont {Hongming}\ \bibnamefont {Weng}}, \bibinfo {author}
  {\bibfnamefont {Xi}~\bibnamefont {Dai}}, \ and\ \bibinfo {author}
  {\bibfnamefont {Zhong}\ \bibnamefont {Fang}},\ }\bibfield  {title} {\enquote
  {\bibinfo {title} {{Dirac semimetal and topological phase transitions in
  A$_3$Bi (A$=$Na, K, Rb)}},}\ }\href {\doibase 10.1103/PhysRevB.85.195320}
  {\bibfield  {journal} {\bibinfo  {journal} {Phys. Rev. B}\ }\textbf {\bibinfo
  {volume} {85}},\ \bibinfo {pages} {195320} (\bibinfo {year}
  {2012})}\BibitemShut {NoStop}%
\bibitem [{\citenamefont {Wang}\ \emph {et~al.}(2013)\citenamefont {Wang},
  \citenamefont {Weng}, \citenamefont {Wu}, \citenamefont {Dai},\ and\
  \citenamefont {Fang}}]{FangCd3As2Dirac}%
  \BibitemOpen
  \bibfield  {author} {\bibinfo {author} {\bibfnamefont {Zhijun}\ \bibnamefont
  {Wang}}, \bibinfo {author} {\bibfnamefont {Hongming}\ \bibnamefont {Weng}},
  \bibinfo {author} {\bibfnamefont {Quansheng}\ \bibnamefont {Wu}}, \bibinfo
  {author} {\bibfnamefont {Xi}~\bibnamefont {Dai}}, \ and\ \bibinfo {author}
  {\bibfnamefont {Zhong}\ \bibnamefont {Fang}},\ }\bibfield  {title} {\enquote
  {\bibinfo {title} {{Three-dimensional Dirac semimetal and quantum transport
  in Cd$_3$As$_2$}},}\ }\href {\doibase 10.1103/PhysRevB.88.125427} {\bibfield
  {journal} {\bibinfo  {journal} {Phys. Rev. B}\ }\textbf {\bibinfo {volume}
  {88}},\ \bibinfo {pages} {125427} (\bibinfo {year} {2013})}\BibitemShut
  {NoStop}%
\bibitem [{Note1()}]{Note1}%
  \BibitemOpen
  \bibinfo {note} {Throughout, we will be careful to distinguish between Dirac
  {\protect \it points} and Weyl {\protect \it nodes}, so that we can refer to
  the two Weyl nodes at a single Dirac point.}\BibitemShut {Stop}%
\bibitem [{\citenamefont {{Liu}}\ \emph {et~al.}(2013)\citenamefont {{Liu}},
  \citenamefont {{Zhou}}, \citenamefont {{Wang}}, \citenamefont {{Weng}},
  \citenamefont {{Prabhakaran}}, \citenamefont {{Mo}}, \citenamefont {{Zhang}},
  \citenamefont {{Shen}}, \citenamefont {{Fang}}, \citenamefont {{Dai}},
  \citenamefont {{Hussain}},\ and\ \citenamefont {{Chen}}}]{Na3BiExpt1}%
  \BibitemOpen
  \bibfield  {author} {\bibinfo {author} {\bibfnamefont {Z.~K.}\ \bibnamefont
  {{Liu}}}, \bibinfo {author} {\bibfnamefont {B.}~\bibnamefont {{Zhou}}},
  \bibinfo {author} {\bibfnamefont {Z.~J.}\ \bibnamefont {{Wang}}}, \bibinfo
  {author} {\bibfnamefont {H.~M.}\ \bibnamefont {{Weng}}}, \bibinfo {author}
  {\bibfnamefont {D.}~\bibnamefont {{Prabhakaran}}}, \bibinfo {author}
  {\bibfnamefont {S.-K.}\ \bibnamefont {{Mo}}}, \bibinfo {author}
  {\bibfnamefont {Y.}~\bibnamefont {{Zhang}}}, \bibinfo {author} {\bibfnamefont
  {Z.~X.}\ \bibnamefont {{Shen}}}, \bibinfo {author} {\bibfnamefont
  {Z.}~\bibnamefont {{Fang}}}, \bibinfo {author} {\bibfnamefont
  {X.}~\bibnamefont {{Dai}}}, \bibinfo {author} {\bibfnamefont
  {Z.}~\bibnamefont {{Hussain}}}, \ and\ \bibinfo {author} {\bibfnamefont
  {Y.~L.}\ \bibnamefont {{Chen}}},\ }\bibfield  {title} {\enquote {\bibinfo
  {title} {{Discovery of a Three-dimensional Topological Dirac Semimetal,
  Na$_3$Bi}},}\ }\href@noop {} {\bibfield  {journal} {\bibinfo  {journal}
  {ArXiv e-prints}\ } (\bibinfo {year} {2013})},\ \Eprint
  {http://arxiv.org/abs/1310.0391} {arXiv:1310.0391 [cond-mat.mtrl-sci]}
  \BibitemShut {NoStop}%
\bibitem [{\citenamefont {{Neupane}}\ \emph {et~al.}(2013)\citenamefont
  {{Neupane}}, \citenamefont {{Xu}}, \citenamefont {{Sankar}}, \citenamefont
  {{Alidoust}}, \citenamefont {{Bian}}, \citenamefont {{Liu}}, \citenamefont
  {{Belopolski}}, \citenamefont {{Chang}}, \citenamefont {{Jeng}},
  \citenamefont {{Lin}}, \citenamefont {{Bansil}}, \citenamefont {{Chou}},\
  and\ \citenamefont {{Hasan}}}]{Cd3As2Expt1}%
  \BibitemOpen
  \bibfield  {author} {\bibinfo {author} {\bibfnamefont {M.}~\bibnamefont
  {{Neupane}}}, \bibinfo {author} {\bibfnamefont {S.}~\bibnamefont {{Xu}}},
  \bibinfo {author} {\bibfnamefont {R.}~\bibnamefont {{Sankar}}}, \bibinfo
  {author} {\bibfnamefont {N.}~\bibnamefont {{Alidoust}}}, \bibinfo {author}
  {\bibfnamefont {G.}~\bibnamefont {{Bian}}}, \bibinfo {author} {\bibfnamefont
  {C.}~\bibnamefont {{Liu}}}, \bibinfo {author} {\bibfnamefont
  {I.}~\bibnamefont {{Belopolski}}}, \bibinfo {author} {\bibfnamefont {T.-R.}\
  \bibnamefont {{Chang}}}, \bibinfo {author} {\bibfnamefont {H.-T.}\
  \bibnamefont {{Jeng}}}, \bibinfo {author} {\bibfnamefont {H.}~\bibnamefont
  {{Lin}}}, \bibinfo {author} {\bibfnamefont {A.}~\bibnamefont {{Bansil}}},
  \bibinfo {author} {\bibfnamefont {F.}~\bibnamefont {{Chou}}}, \ and\ \bibinfo
  {author} {\bibfnamefont {M.~Z.}\ \bibnamefont {{Hasan}}},\ }\bibfield
  {title} {\enquote {\bibinfo {title} {{Observation of a topological 3D Dirac
  semimetal phase in high-mobility Cd3As2 and related materials}},}\
  }\href@noop {} {\bibfield  {journal} {\bibinfo  {journal} {ArXiv e-prints}\ }
  (\bibinfo {year} {2013})},\ \Eprint {http://arxiv.org/abs/1309.7892}
  {arXiv:1309.7892 [cond-mat.mes-hall]} \BibitemShut {NoStop}%
\bibitem [{\citenamefont {{Borisenko}}\ \emph {et~al.}(2013)\citenamefont
  {{Borisenko}}, \citenamefont {{Gibson}}, \citenamefont {{Evtushinsky}},
  \citenamefont {{Zabolotnyy}}, \citenamefont {{Buechner}},\ and\ \citenamefont
  {{Cava}}}]{Cd3As2Expt2}%
  \BibitemOpen
  \bibfield  {author} {\bibinfo {author} {\bibfnamefont {S.}~\bibnamefont
  {{Borisenko}}}, \bibinfo {author} {\bibfnamefont {Q.}~\bibnamefont
  {{Gibson}}}, \bibinfo {author} {\bibfnamefont {D.}~\bibnamefont
  {{Evtushinsky}}}, \bibinfo {author} {\bibfnamefont {V.}~\bibnamefont
  {{Zabolotnyy}}}, \bibinfo {author} {\bibfnamefont {B.}~\bibnamefont
  {{Buechner}}}, \ and\ \bibinfo {author} {\bibfnamefont {R.~J.}\ \bibnamefont
  {{Cava}}},\ }\bibfield  {title} {\enquote {\bibinfo {title} {{Experimental
  Realization of a Three-Dimensional Dirac Semimetal}},}\ }\href@noop {}
  {\bibfield  {journal} {\bibinfo  {journal} {ArXiv e-prints}\ } (\bibinfo
  {year} {2013})},\ \Eprint {http://arxiv.org/abs/1309.7978} {arXiv:1309.7978
  [cond-mat.mes-hall]} \BibitemShut {NoStop}%
\bibitem [{\citenamefont {{Xu}}\ \emph {et~al.}(2013)\citenamefont {{Xu}},
  \citenamefont {{Liu}}, \citenamefont {{Kushwaha}}, \citenamefont {{Chang}},
  \citenamefont {{Krizan}}, \citenamefont {{Sankar}}, \citenamefont {{Polley}},
  \citenamefont {{Adell}}, \citenamefont {{Balasubramanian}}, \citenamefont
  {{Miyamoto}}, \citenamefont {{Alidoust}}, \citenamefont {{Bian}},
  \citenamefont {{Neupane}}, \citenamefont {{Belopolski}}, \citenamefont
  {{Jeng}}, \citenamefont {{Huang}}, \citenamefont {{Tsai}}, \citenamefont
  {{Lin}}, \citenamefont {{Chou}}, \citenamefont {{Okuda}}, \citenamefont
  {{Bansil}}, \citenamefont {{Cava}},\ and\ \citenamefont
  {{Hasan}}}]{Na3BiExpt2}%
  \BibitemOpen
  \bibfield  {author} {\bibinfo {author} {\bibfnamefont {S.-Y.}\ \bibnamefont
  {{Xu}}}, \bibinfo {author} {\bibfnamefont {C.}~\bibnamefont {{Liu}}},
  \bibinfo {author} {\bibfnamefont {S.~K.}\ \bibnamefont {{Kushwaha}}},
  \bibinfo {author} {\bibfnamefont {T.-R.}\ \bibnamefont {{Chang}}}, \bibinfo
  {author} {\bibfnamefont {J.~W.}\ \bibnamefont {{Krizan}}}, \bibinfo {author}
  {\bibfnamefont {R.}~\bibnamefont {{Sankar}}}, \bibinfo {author}
  {\bibfnamefont {C.~M.}\ \bibnamefont {{Polley}}}, \bibinfo {author}
  {\bibfnamefont {J.}~\bibnamefont {{Adell}}}, \bibinfo {author} {\bibfnamefont
  {T.}~\bibnamefont {{Balasubramanian}}}, \bibinfo {author} {\bibfnamefont
  {K.}~\bibnamefont {{Miyamoto}}}, \bibinfo {author} {\bibfnamefont
  {N.}~\bibnamefont {{Alidoust}}}, \bibinfo {author} {\bibfnamefont
  {G.}~\bibnamefont {{Bian}}}, \bibinfo {author} {\bibfnamefont
  {M.}~\bibnamefont {{Neupane}}}, \bibinfo {author} {\bibfnamefont
  {I.}~\bibnamefont {{Belopolski}}}, \bibinfo {author} {\bibfnamefont {H.-T.}\
  \bibnamefont {{Jeng}}}, \bibinfo {author} {\bibfnamefont {C.-Y.}\
  \bibnamefont {{Huang}}}, \bibinfo {author} {\bibfnamefont {W.-F.}\
  \bibnamefont {{Tsai}}}, \bibinfo {author} {\bibfnamefont {H.}~\bibnamefont
  {{Lin}}}, \bibinfo {author} {\bibfnamefont {F.~C.}\ \bibnamefont {{Chou}}},
  \bibinfo {author} {\bibfnamefont {T.}~\bibnamefont {{Okuda}}}, \bibinfo
  {author} {\bibfnamefont {A.}~\bibnamefont {{Bansil}}}, \bibinfo {author}
  {\bibfnamefont {R.~J.}\ \bibnamefont {{Cava}}}, \ and\ \bibinfo {author}
  {\bibfnamefont {M.~Z.}\ \bibnamefont {{Hasan}}},\ }\bibfield  {title}
  {\enquote {\bibinfo {title} {{Observation of a bulk 3D Dirac multiplet,
  Lifshitz transition, and nestled spin states in Na3Bi}},}\ }\href@noop {}
  {\bibfield  {journal} {\bibinfo  {journal} {ArXiv e-prints}\ } (\bibinfo
  {year} {2013})},\ \Eprint {http://arxiv.org/abs/1312.7624} {arXiv:1312.7624
  [cond-mat.mes-hall]} \BibitemShut {NoStop}%
\bibitem [{\citenamefont {Ong}(2013)}]{Ong_unpub}%
  \BibitemOpen
  \bibfield  {author} {\bibinfo {author} {\bibfnamefont {N.~P.}\ \bibnamefont
  {Ong}},\ }\href@noop {} {\enquote {\bibinfo {title} {private
  communication},}\ } (\bibinfo {year} {2013})\BibitemShut {NoStop}%
\bibitem [{\citenamefont {Adler}(1969)}]{AdlerAnomaly}%
  \BibitemOpen
  \bibfield  {author} {\bibinfo {author} {\bibfnamefont {Stephen~L.}\
  \bibnamefont {Adler}},\ }\bibfield  {title} {\enquote {\bibinfo {title}
  {Axial-vector vertex in spinor electrodynamics},}\ }\href {\doibase
  10.1103/PhysRev.177.2426} {\bibfield  {journal} {\bibinfo  {journal} {Phys.
  Rev.}\ }\textbf {\bibinfo {volume} {177}},\ \bibinfo {pages} {2426--2438}
  (\bibinfo {year} {1969})}\BibitemShut {NoStop}%
\bibitem [{\citenamefont {Bell}\ and\ \citenamefont
  {Jackiw}(1969)}]{BellJackiwAnomaly}%
  \BibitemOpen
  \bibfield  {author} {\bibinfo {author} {\bibfnamefont {J.~S.}\ \bibnamefont
  {Bell}}\ and\ \bibinfo {author} {\bibfnamefont {R.}~\bibnamefont {Jackiw}},\
  }\bibfield  {title} {\enquote {\bibinfo {title} {{A PCAC puzzle: $\pi_0
  \rightarrow \gamma\gamma$ in the $\sigma$-model}},}\ }\href {\doibase
  10.1007/BF02823296} {\bibfield  {journal} {\bibinfo  {journal} {Il Nuovo
  Cimento A}\ }\textbf {\bibinfo {volume} {60}},\ \bibinfo {pages} {47--61}
  (\bibinfo {year} {1969})}\BibitemShut {NoStop}%
\bibitem [{\citenamefont {Abanin}\ \emph {et~al.}(2009)\citenamefont {Abanin},
  \citenamefont {Shytov}, \citenamefont {Levitov},\ and\ \citenamefont
  {Halperin}}]{AbaninNonlocal}%
  \BibitemOpen
  \bibfield  {author} {\bibinfo {author} {\bibfnamefont {D.~A.}\ \bibnamefont
  {Abanin}}, \bibinfo {author} {\bibfnamefont {A.~V.}\ \bibnamefont {Shytov}},
  \bibinfo {author} {\bibfnamefont {L.~S.}\ \bibnamefont {Levitov}}, \ and\
  \bibinfo {author} {\bibfnamefont {B.~I.}\ \bibnamefont {Halperin}},\
  }\bibfield  {title} {\enquote {\bibinfo {title} {Nonlocal charge transport
  mediated by spin diffusion in the spin hall effect regime},}\ }\href
  {\doibase 10.1103/PhysRevB.79.035304} {\bibfield  {journal} {\bibinfo
  {journal} {Phys. Rev. B}\ }\textbf {\bibinfo {volume} {79}},\ \bibinfo
  {pages} {035304} (\bibinfo {year} {2009})}\BibitemShut {NoStop}%
\bibitem [{\citenamefont {Abanin}\ \emph
  {et~al.}(2011{\natexlab{a}})\citenamefont {Abanin}, \citenamefont {Morozov},
  \citenamefont {Ponomarenko}, \citenamefont {Gorbachev}, \citenamefont
  {Mayorov}, \citenamefont {Katsnelson}, \citenamefont {Watanabe},
  \citenamefont {Taniguchi}, \citenamefont {Novoselov}, \citenamefont
  {Levitov},\ and\ \citenamefont {Geim}}]{AbaninGiantNonlocality}%
  \BibitemOpen
  \bibfield  {author} {\bibinfo {author} {\bibfnamefont {D.~A.}\ \bibnamefont
  {Abanin}}, \bibinfo {author} {\bibfnamefont {S.~V.}\ \bibnamefont {Morozov}},
  \bibinfo {author} {\bibfnamefont {L.~A.}\ \bibnamefont {Ponomarenko}},
  \bibinfo {author} {\bibfnamefont {R.~V.}\ \bibnamefont {Gorbachev}}, \bibinfo
  {author} {\bibfnamefont {A.~S.}\ \bibnamefont {Mayorov}}, \bibinfo {author}
  {\bibfnamefont {M.~I.}\ \bibnamefont {Katsnelson}}, \bibinfo {author}
  {\bibfnamefont {K.}~\bibnamefont {Watanabe}}, \bibinfo {author}
  {\bibfnamefont {T.}~\bibnamefont {Taniguchi}}, \bibinfo {author}
  {\bibfnamefont {K.~S.}\ \bibnamefont {Novoselov}}, \bibinfo {author}
  {\bibfnamefont {L.~S.}\ \bibnamefont {Levitov}}, \ and\ \bibinfo {author}
  {\bibfnamefont {A.~K.}\ \bibnamefont {Geim}},\ }\bibfield  {title} {\enquote
  {\bibinfo {title} {{Giant Nonlocality Near the Dirac Point in Graphene}},}\
  }\href {\doibase 10.1126/science.1199595} {\bibfield  {journal} {\bibinfo
  {journal} {Science}\ }\textbf {\bibinfo {volume} {332}},\ \bibinfo {pages}
  {328--330} (\bibinfo {year} {2011}{\natexlab{a}})}\BibitemShut {NoStop}%
\bibitem [{\citenamefont {Abanin}\ \emph
  {et~al.}(2011{\natexlab{b}})\citenamefont {Abanin}, \citenamefont
  {Gorbachev}, \citenamefont {Novoselov}, \citenamefont {Geim},\ and\
  \citenamefont {Levitov}}]{AbaninZeemanSHE}%
  \BibitemOpen
  \bibfield  {author} {\bibinfo {author} {\bibfnamefont {D.~A.}\ \bibnamefont
  {Abanin}}, \bibinfo {author} {\bibfnamefont {R.~V.}\ \bibnamefont
  {Gorbachev}}, \bibinfo {author} {\bibfnamefont {K.~S.}\ \bibnamefont
  {Novoselov}}, \bibinfo {author} {\bibfnamefont {A.~K.}\ \bibnamefont {Geim}},
  \ and\ \bibinfo {author} {\bibfnamefont {L.~S.}\ \bibnamefont {Levitov}},\
  }\bibfield  {title} {\enquote {\bibinfo {title} {{Giant Spin-Hall Effect
  Induced by the Zeeman Interaction in Graphene}},}\ }\href {\doibase
  10.1103/PhysRevLett.107.096601} {\bibfield  {journal} {\bibinfo  {journal}
  {Phys. Rev. Lett.}\ }\textbf {\bibinfo {volume} {107}},\ \bibinfo {pages}
  {096601} (\bibinfo {year} {2011}{\natexlab{b}})}\BibitemShut {NoStop}%
\bibitem [{\citenamefont {Balakrishnan}\ \emph {et~al.}(2013)\citenamefont
  {Balakrishnan}, \citenamefont {Kok Wai~Koon}, \citenamefont {Jaiswal},
  \citenamefont {Castro~Neto},\ and\ \citenamefont
  {Ozyilmaz}}]{Balakrishnan:2013fk}%
  \BibitemOpen
  \bibfield  {author} {\bibinfo {author} {\bibfnamefont {Jayakumar}\
  \bibnamefont {Balakrishnan}}, \bibinfo {author} {\bibfnamefont {Gavin}\
  \bibnamefont {Kok Wai~Koon}}, \bibinfo {author} {\bibfnamefont {Manu}\
  \bibnamefont {Jaiswal}}, \bibinfo {author} {\bibfnamefont {A.~H.}\
  \bibnamefont {Castro~Neto}}, \ and\ \bibinfo {author} {\bibfnamefont
  {Barbaros}\ \bibnamefont {Ozyilmaz}},\ }\bibfield  {title} {\enquote
  {\bibinfo {title} {Colossal enhancement of spin-orbit coupling in weakly
  hydrogenated graphene},}\ }\href {http://dx.doi.org/10.1038/nphys2576}
  {\bibfield  {journal} {\bibinfo  {journal} {Nat Phys}\ }\textbf {\bibinfo
  {volume} {9}},\ \bibinfo {pages} {284--287} (\bibinfo {year}
  {2013})}\BibitemShut {NoStop}%
\bibitem [{\citenamefont {Aji}(2012)}]{AjiABJAnomaly}%
  \BibitemOpen
  \bibfield  {author} {\bibinfo {author} {\bibfnamefont {Vivek}\ \bibnamefont
  {Aji}},\ }\bibfield  {title} {\enquote {\bibinfo {title} {{Adler-Bell-Jackiw
  anomaly in Weyl semimetals: Application to pyrochlore iridates}},}\ }\href
  {\doibase 10.1103/PhysRevB.85.241101} {\bibfield  {journal} {\bibinfo
  {journal} {Phys. Rev. B}\ }\textbf {\bibinfo {volume} {85}},\ \bibinfo
  {pages} {241101} (\bibinfo {year} {2012})}\BibitemShut {NoStop}%
\bibitem [{\citenamefont {{Son}}\ and\ \citenamefont
  {{Spivak}}(2012)}]{SonSpivak}%
  \BibitemOpen
  \bibfield  {author} {\bibinfo {author} {\bibfnamefont {D.~T.}\ \bibnamefont
  {{Son}}}\ and\ \bibinfo {author} {\bibfnamefont {B.~Z.}\ \bibnamefont
  {{Spivak}}},\ }\bibfield  {title} {\enquote {\bibinfo {title} {{Chiral
  Anomaly and Classical Negative Magnetoresistance of Weyl Metals}},}\
  }\href@noop {} {\bibfield  {journal} {\bibinfo  {journal} {ArXiv e-prints}\ }
  (\bibinfo {year} {2012})},\ \Eprint {http://arxiv.org/abs/1206.1627}
  {arXiv:1206.1627 [cond-mat.mes-hall]} \BibitemShut {NoStop}%
\bibitem [{\citenamefont {Burkov}\ \emph {et~al.}(2011)\citenamefont {Burkov},
  \citenamefont {Hook},\ and\ \citenamefont {Balents}}]{BurkovTopoNodal}%
  \BibitemOpen
  \bibfield  {author} {\bibinfo {author} {\bibfnamefont {A.~A.}\ \bibnamefont
  {Burkov}}, \bibinfo {author} {\bibfnamefont {M.~D.}\ \bibnamefont {Hook}}, \
  and\ \bibinfo {author} {\bibfnamefont {Leon}\ \bibnamefont {Balents}},\
  }\bibfield  {title} {\enquote {\bibinfo {title} {Topological nodal
  semimetals},}\ }\href {\doibase 10.1103/PhysRevB.84.235126} {\bibfield
  {journal} {\bibinfo  {journal} {Phys. Rev. B}\ }\textbf {\bibinfo {volume}
  {84}},\ \bibinfo {pages} {235126} (\bibinfo {year} {2011})}\BibitemShut
  {NoStop}%
\bibitem [{\citenamefont {{Liu}}\ \emph {et~al.}(2012)\citenamefont {{Liu}},
  \citenamefont {{Ye}},\ and\ \citenamefont {{Qi}}}]{QiPlasmons}%
  \BibitemOpen
  \bibfield  {author} {\bibinfo {author} {\bibfnamefont {C.-X.}\ \bibnamefont
  {{Liu}}}, \bibinfo {author} {\bibfnamefont {P.}~\bibnamefont {{Ye}}}, \ and\
  \bibinfo {author} {\bibfnamefont {X.-L.}\ \bibnamefont {{Qi}}},\ }\bibfield
  {title} {\enquote {\bibinfo {title} {{Chiral gauge field and axial anomaly in
  a Weyl semi-metal}},}\ }\href@noop {} {\bibfield  {journal} {\bibinfo
  {journal} {ArXiv e-prints}\ } (\bibinfo {year} {2012})},\ \Eprint
  {http://arxiv.org/abs/1204.6551} {arXiv:1204.6551 [cond-mat.str-el]}
  \BibitemShut {NoStop}%
\bibitem [{\citenamefont {Yang}\ \emph {et~al.}(2011)\citenamefont {Yang},
  \citenamefont {Lu},\ and\ \citenamefont {Ran}}]{RanWeyl}%
  \BibitemOpen
  \bibfield  {author} {\bibinfo {author} {\bibfnamefont {Kai-Yu}\ \bibnamefont
  {Yang}}, \bibinfo {author} {\bibfnamefont {Yuan-Ming}\ \bibnamefont {Lu}}, \
  and\ \bibinfo {author} {\bibfnamefont {Ying}\ \bibnamefont {Ran}},\
  }\bibfield  {title} {\enquote {\bibinfo {title} {{Quantum Hall effects in a
  Weyl semimetal: Possible application in pyrochlore iridates}},}\ }\href
  {\doibase 10.1103/PhysRevB.84.075129} {\bibfield  {journal} {\bibinfo
  {journal} {Phys. Rev. B}\ }\textbf {\bibinfo {volume} {84}},\ \bibinfo
  {pages} {075129} (\bibinfo {year} {2011})}\BibitemShut {NoStop}%
\bibitem [{Note2()}]{Note2}%
  \BibitemOpen
  \bibinfo {note} {For instance, for $B\not =0$ there is always a net uniform
  valley current $\protect \boldsymbol {j}^R-\protect \boldsymbol {j}^L \propto
  \protect \mathbf {B} $ even in equilibrium, but this is unimportant to the
  transport calculation and we therefore ignore it.}\BibitemShut {Stop}%
\bibitem [{Note3()}]{Note3}%
  \BibitemOpen
  \bibinfo {note} {Throughout we assume an isotropic node
  dispersion.}\BibitemShut {Stop}%
\bibitem [{\citenamefont {Son}\ and\ \citenamefont
  {Sur\'owka}(2009)}]{sonsurowka}%
  \BibitemOpen
  \bibfield  {author} {\bibinfo {author} {\bibfnamefont {Dam~T.}\ \bibnamefont
  {Son}}\ and\ \bibinfo {author} {\bibfnamefont {Piotr}\ \bibnamefont
  {Sur\'owka}},\ }\bibfield  {title} {\enquote {\bibinfo {title} {Hydrodynamics
  with triangle anomalies},}\ }\href {\doibase 10.1103/PhysRevLett.103.191601}
  {\bibfield  {journal} {\bibinfo  {journal} {Phys. Rev. Lett.}\ }\textbf
  {\bibinfo {volume} {103}},\ \bibinfo {pages} {191601} (\bibinfo {year}
  {2009})}\BibitemShut {NoStop}%
\bibitem [{Note4()}]{Note4}%
  \BibitemOpen
  \bibinfo {note} {Note the extra factor of $e$ for the electrical
  current.}\BibitemShut {Stop}%
\bibitem [{Note5()}]{Note5}%
  \BibitemOpen
  \bibinfo {note} {Although not crucial, this simplifies the
  analysis.}\BibitemShut {Stop}%
\bibitem [{Note6()}]{Note6}%
  \BibitemOpen
  \bibinfo {note} {The precise number will depend on the sample
  size.}\BibitemShut {Stop}%
\bibitem [{Note7()}]{Note7}%
  \BibitemOpen
  \bibinfo {note} {A similar problem arises with spin injection with
  ferromagnetic leads.}\BibitemShut {Stop}%
\bibitem [{\citenamefont {Rashba}(2000)}]{Rashba2000mismatch}%
  \BibitemOpen
  \bibfield  {author} {\bibinfo {author} {\bibfnamefont {E.~I.}\ \bibnamefont
  {Rashba}},\ }\bibfield  {title} {\enquote {\bibinfo {title} {Theory of
  electrical spin injection: Tunnel contacts as a solution of the conductivity
  mismatch problem},}\ }\href {\doibase 10.1103/PhysRevB.62.R16267} {\bibfield
  {journal} {\bibinfo  {journal} {Phys. Rev. B}\ }\textbf {\bibinfo {volume}
  {62}},\ \bibinfo {pages} {R16267--R16270} (\bibinfo {year}
  {2000})}\BibitemShut {NoStop}%
\bibitem [{Note8()}]{Note8}%
  \BibitemOpen
  \bibinfo {note} {In the opposite limit, even a spherically symmetric impurity
  might lead to mixing between isospins, as the placement of the impurity away
  from the symmetry center would strongly break the crystalline point-group
  symmetry.}\BibitemShut {Stop}%
\bibitem [{\citenamefont {Parameswaran}\ \emph {et~al.}(2014)\citenamefont
  {Parameswaran}, \citenamefont {Pesin}, \citenamefont {Abanin},\ and\
  \citenamefont {Vishwanath}}]{PPAV-unpub}%
  \BibitemOpen
  \bibfield  {author} {\bibinfo {author} {\bibfnamefont {S.~A.}\ \bibnamefont
  {Parameswaran}}, \bibinfo {author} {\bibfnamefont {D.~A.}\ \bibnamefont
  {Pesin}}, \bibinfo {author} {\bibfnamefont {D.~A.}\ \bibnamefont {Abanin}}, \
  and\ \bibinfo {author} {\bibfnamefont {Ashvin}\ \bibnamefont {Vishwanath}},\
  }\href@noop {} {\enquote {\bibinfo {title} {{Impurity Scattering and
  Chirality Relaxation in 3D Dirac Semimetals}},}\ } (\bibinfo {year}
  {2014})\BibitemShut {NoStop}%
\bibitem [{Note9()}]{Note9}%
  \BibitemOpen
  \bibinfo {note} {Note that for simplicity we have ignored an
  inversion-breaking term that is required to properly describe Cadmium
  Arsenide; we defer a discussion of such details to future work.}\BibitemShut
  {Stop}%
\bibitem [{Note10()}]{Note10}%
  \BibitemOpen
  \bibinfo {note} {Observe that in contrast to the case of vacuum Dirac
  fermions familiar to high-energy physicists, this decomposition is
  unambiguous in a crystal since electrons emanating from the two Weyl points
  carry distinct crystal symmetry quantum numbers; this is also why a mass term
  is forbidden by symmetry.}\BibitemShut {Stop}%
\bibitem [{\citenamefont {Yanagishima}\ and\ \citenamefont
  {Maeno}(2001)}]{WeylResistivityMaeno}%
  \BibitemOpen
  \bibfield  {author} {\bibinfo {author} {\bibfnamefont {Daiki}\ \bibnamefont
  {Yanagishima}}\ and\ \bibinfo {author} {\bibfnamefont {Yoshiteru}\
  \bibnamefont {Maeno}},\ }\bibfield  {title} {\enquote {\bibinfo {title}
  {Metal-nonmetal changeover in pyrochlore iridates},}\ }\href {\doibase
  10.1143/JPSJ.70.2880} {\bibfield  {journal} {\bibinfo  {journal} {Journal of
  the Physical Society of Japan}\ }\textbf {\bibinfo {volume} {70}},\ \bibinfo
  {pages} {2880--2883} (\bibinfo {year} {2001})}\BibitemShut {NoStop}%
\bibitem [{\citenamefont {Tafti}\ \emph {et~al.}(2012)\citenamefont {Tafti},
  \citenamefont {Ishikawa}, \citenamefont {McCollam}, \citenamefont
  {Nakatsuji},\ and\ \citenamefont {Julian}}]{EuIridateExperiments}%
  \BibitemOpen
  \bibfield  {author} {\bibinfo {author} {\bibfnamefont {F.~F.}\ \bibnamefont
  {Tafti}}, \bibinfo {author} {\bibfnamefont {J.~J.}\ \bibnamefont {Ishikawa}},
  \bibinfo {author} {\bibfnamefont {A.}~\bibnamefont {McCollam}}, \bibinfo
  {author} {\bibfnamefont {S.}~\bibnamefont {Nakatsuji}}, \ and\ \bibinfo
  {author} {\bibfnamefont {S.~R.}\ \bibnamefont {Julian}},\ }\bibfield  {title}
  {\enquote {\bibinfo {title} {{Pressure-tuned insulator to metal transition in
  ${\mathbf{Eu}}_{\mathbf{2}}{\mathbf{Ir}}_{\mathbf{2}}{\mathbf{O}}_{\mathbf{7}}$}},}\
  }\href {\doibase 10.1103/PhysRevB.85.205104} {\bibfield  {journal} {\bibinfo
  {journal} {Phys. Rev. B}\ }\textbf {\bibinfo {volume} {85}},\ \bibinfo
  {pages} {205104} (\bibinfo {year} {2012})}\BibitemShut {NoStop}%
\bibitem [{\citenamefont {Xu}\ \emph {et~al.}(2011)\citenamefont {Xu},
  \citenamefont {Weng}, \citenamefont {Wang}, \citenamefont {Dai},\ and\
  \citenamefont {Fang}}]{XuWSM-HgCrSe}%
  \BibitemOpen
  \bibfield  {author} {\bibinfo {author} {\bibfnamefont {Gang}\ \bibnamefont
  {Xu}}, \bibinfo {author} {\bibfnamefont {Hongming}\ \bibnamefont {Weng}},
  \bibinfo {author} {\bibfnamefont {Zhijun}\ \bibnamefont {Wang}}, \bibinfo
  {author} {\bibfnamefont {Xi}~\bibnamefont {Dai}}, \ and\ \bibinfo {author}
  {\bibfnamefont {Zhong}\ \bibnamefont {Fang}},\ }\bibfield  {title} {\enquote
  {\bibinfo {title} {{Chern Semimetal and the Quantized Anomalous Hall Effect
  in ${\mathrm{HgCr}}_{2}{\mathrm{Se}}_{4}$}},}\ }\href {\doibase
  10.1103/PhysRevLett.107.186806} {\bibfield  {journal} {\bibinfo  {journal}
  {Phys. Rev. Lett.}\ }\textbf {\bibinfo {volume} {107}},\ \bibinfo {pages}
  {186806} (\bibinfo {year} {2011})}\BibitemShut {NoStop}%
\bibitem [{\citenamefont {Witczak-Krempa}\ and\ \citenamefont
  {Kim}(2012)}]{Kim}%
  \BibitemOpen
  \bibfield  {author} {\bibinfo {author} {\bibfnamefont {William}\ \bibnamefont
  {Witczak-Krempa}}\ and\ \bibinfo {author} {\bibfnamefont {Yong~Baek}\
  \bibnamefont {Kim}},\ }\bibfield  {title} {\enquote {\bibinfo {title}
  {Topological and magnetic phases of interacting electrons in the pyrochlore
  iridates},}\ }\href {\doibase 10.1103/PhysRevB.85.045124} {\bibfield
  {journal} {\bibinfo  {journal} {Phys. Rev. B}\ }\textbf {\bibinfo {volume}
  {85}},\ \bibinfo {pages} {045124} (\bibinfo {year} {2012})}\BibitemShut
  {NoStop}%
\bibitem [{\citenamefont {Ma\~nes}(2012)}]{Manes}%
  \BibitemOpen
  \bibfield  {author} {\bibinfo {author} {\bibfnamefont {J.~L.}\ \bibnamefont
  {Ma\~nes}},\ }\bibfield  {title} {\enquote {\bibinfo {title} {Existence of
  bulk chiral fermions and crystal symmetry},}\ }\href {\doibase
  10.1103/PhysRevB.85.155118} {\bibfield  {journal} {\bibinfo  {journal} {Phys.
  Rev. B}\ }\textbf {\bibinfo {volume} {85}},\ \bibinfo {pages} {155118}
  (\bibinfo {year} {2012})}\BibitemShut {NoStop}%
\bibitem [{\citenamefont {Wan}\ \emph {et~al.}(2012)\citenamefont {Wan},
  \citenamefont {Vishwanath},\ and\ \citenamefont {Savrasov}}]{WanOsmates}%
  \BibitemOpen
  \bibfield  {author} {\bibinfo {author} {\bibfnamefont {Xiangang}\
  \bibnamefont {Wan}}, \bibinfo {author} {\bibfnamefont {Ashvin}\ \bibnamefont
  {Vishwanath}}, \ and\ \bibinfo {author} {\bibfnamefont {Sergey~Y.}\
  \bibnamefont {Savrasov}},\ }\bibfield  {title} {\enquote {\bibinfo {title}
  {Computational design of axion insulators based on $5d$ spinel compounds},}\
  }\href {\doibase 10.1103/PhysRevLett.108.146601} {\bibfield  {journal}
  {\bibinfo  {journal} {Phys. Rev. Lett.}\ }\textbf {\bibinfo {volume} {108}},\
  \bibinfo {pages} {146601} (\bibinfo {year} {2012})}\BibitemShut {NoStop}%
\bibitem [{\citenamefont {{Chen}}\ \emph {et~al.}(2013)\citenamefont {{Chen}},
  \citenamefont {{Bergman}},\ and\ \citenamefont {{Burkov}}}]{WeylAHEMFM}%
  \BibitemOpen
  \bibfield  {author} {\bibinfo {author} {\bibfnamefont {Y.}~\bibnamefont
  {{Chen}}}, \bibinfo {author} {\bibfnamefont {D.~L.}\ \bibnamefont
  {{Bergman}}}, \ and\ \bibinfo {author} {\bibfnamefont {A.~A.}\ \bibnamefont
  {{Burkov}}},\ }\bibfield  {title} {\enquote {\bibinfo {title} {{Weyl Fermions
  and the Anomalous Hall Effect in Metallic Ferromagnets}},}\ }\href@noop {}
  {\bibfield  {journal} {\bibinfo  {journal} {ArXiv e-prints}\ } (\bibinfo
  {year} {2013})},\ \Eprint {http://arxiv.org/abs/1305.0183} {arXiv:1305.0183
  [cond-mat.mes-hall]} \BibitemShut {NoStop}%
\bibitem [{\citenamefont {Burkov}\ and\ \citenamefont
  {Balents}(2011)}]{WeylMultiLayer}%
  \BibitemOpen
  \bibfield  {author} {\bibinfo {author} {\bibfnamefont {A.~A.}\ \bibnamefont
  {Burkov}}\ and\ \bibinfo {author} {\bibfnamefont {Leon}\ \bibnamefont
  {Balents}},\ }\bibfield  {title} {\enquote {\bibinfo {title} {Weyl semimetal
  in a topological insulator multilayer},}\ }\href {\doibase
  10.1103/PhysRevLett.107.127205} {\bibfield  {journal} {\bibinfo  {journal}
  {Phys. Rev. Lett.}\ }\textbf {\bibinfo {volume} {107}},\ \bibinfo {pages}
  {127205} (\bibinfo {year} {2011})}\BibitemShut {NoStop}%
\bibitem [{\citenamefont {Hal\'asz}\ and\ \citenamefont
  {Balents}(2012)}]{HalaszBalents}%
  \BibitemOpen
  \bibfield  {author} {\bibinfo {author} {\bibfnamefont {G\'abor~B.}\
  \bibnamefont {Hal\'asz}}\ and\ \bibinfo {author} {\bibfnamefont {Leon}\
  \bibnamefont {Balents}},\ }\bibfield  {title} {\enquote {\bibinfo {title}
  {{Time-reversal invariant realization of the Weyl semimetal phase}},}\ }\href
  {\doibase 10.1103/PhysRevB.85.035103} {\bibfield  {journal} {\bibinfo
  {journal} {Phys. Rev. B}\ }\textbf {\bibinfo {volume} {85}},\ \bibinfo
  {pages} {035103} (\bibinfo {year} {2012})}\BibitemShut {NoStop}%
\bibitem [{\citenamefont {{Cho}}(2011)}]{Cho}%
  \BibitemOpen
  \bibfield  {author} {\bibinfo {author} {\bibfnamefont {G.~Y.}\ \bibnamefont
  {{Cho}}},\ }\bibfield  {title} {\enquote {\bibinfo {title} {{Possible
  topological phases of bulk magnetically doped Bi$_2$Se$_3$: turning a
  topological band insulator into the Weyl semimetal}},}\ }\href@noop {}
  {\bibfield  {journal} {\bibinfo  {journal} {ArXiv e-prints}\ } (\bibinfo
  {year} {2011})},\ \Eprint {http://arxiv.org/abs/1110.1939} {arXiv:1110.1939
  [cond-mat.str-el]} \BibitemShut {NoStop}%
\end{thebibliography}%

\end{document}